\documentclass[twocolumn]{aastex701}

\usepackage{amsmath}
\usepackage{placeins}
\usepackage{hyperref}

\newcommand\earthmass{M_{\oplus}}
\newcommand\earthradius{R_{\oplus}}
\newcommand\solmass{M_{\odot}}
\newcommand\solradius{R_{\odot}}
\newcommand\sollum{L_{\odot}}

\begin{document}

\title{\textit{Searching for GEMS:} TOI-5916 b \& TOI-6158 b are two Saturn-density planets orbiting M2 dwarfs}

\author[0009-0005-9015-4699]{Shane O'Brien}
\affiliation{Department of Physics \& Astronomy, University of California, Irvine, CA 92697, USA}
\email{sgobrien@uci.edu}

\author[0009-0005-6571-4818]{Amber Wong}
\affiliation{Department of Physics \& Astronomy, University of California, Irvine, CA 92697, USA}
\email{amberyw@uci.edu}

\author[0000-0002-7127-7643]{Te Han}
\affiliation{Department of Physics \& Astronomy, University of California, Irvine, CA 92697, USA}
\email{teh2@uci.edu}

\author[0000-0003-0149-9678]{Paul Robertson}
\affiliation{Department of Physics \& Astronomy, University of California, Irvine, CA 92697, USA}
\email{paul.robertson@uci.edu}

\author[0000-0001-8401-4300]{Shubham Kanodia}
\affiliation{Earth and Planets Laboratory, Carnegie Institution for Science, 5241 Broad Branch Road, NW, Washington, DC 20015, USA}
\email{skanodia@carnegiescience.edu}

\author[0000-0003-4835-0619]{Caleb I. Ca\~nas}
\altaffiliation{NASA Postdoctoral Program Fellow}
\affiliation{NASA Goddard Space Flight Center, 8800 Greenbelt Road, Greenbelt, MD 20771, USA}
\email{c.canas@nasa.gov}

\author[0000-0002-5463-9980]{Arvind F. Gupta}
\affiliation{U.S. National Science Foundation, National Optical-Infrared Astronomy Research Laboratory, 950 N. Cherry Ave., Tucson, AZ 85719, USA}
\email{arvind.gupta@noirlab.edu}


\author[0000-0002-5817-202X]{Tera Swaby}
\affiliation{Department of Physics \& Astronomy, University of Wyoming, Laramie, WY 82070, USA}
\email{tswaby@uwyo.edu}

\author[0000-0002-4475-4176]{Henry A. Kobulnicky}
\affiliation{Department of Physics \& Astronomy, University of Wyoming, Laramie, WY 82070, USA}
\email{chipk@uwyo.edu}

\author[0000-0003-2535-3091]{Nidia Morrell}
\affiliation{Las Campanas Observatory, Carnegie Observatories, Casilla 601, La Serena, Chile}
\email{nmorrell@carnegiescience.edu}

\author[0009-0009-4977-1010]{Michael Rodruck} 
\affiliation{Department of Physics, Engineering, and Astrophysics, Randolph-Macon College, Ashland, VA 23005, USA}
\email{mrodruck@gmail.com}

\author[0000-0002-9082-6337]{Andrea S.J.\ Lin}
\affiliation{\Caltech}
\email{asjlin@caltech.edu}

\author[0000-0002-0048-2586]{Andrew Monson}
\affiliation{\UA}
\email{monson.andy@gmail.com}

\author[0000-0001-9662-3496]{William D. Cochran}
\affiliation{McDonald Observatory and Center for Planetary Systems Habitability, The University of Texas, Austin TX 78712 USA.}
\email{wdc@astro.as.utexas.edu}

\author[0000-0003-4384-7220]{Chad F.\ Bender}
\affiliation{\UA}
\email{cbender@arizona.edu}

\author[0000-0002-2144-0764]{Scott A.\ Diddams}
\affiliation{\CUBoulderEE}
\affiliation{\CUBoulder}
\email{scott.diddams@colorado.edu}

\author[0000-0003-1312-9391]{Samuel Halverson}
\affiliation{\JPL}
\email{samuel.halverson@jpl.nasa.gov}

\author[0000-0001-9626-0613]{Daniel M.\ Krolikowski}
\affiliation{\UA}
\email{krolikowski@arizona.edu}

\author[0000-0002-2990-7613]{Jessica E.~Libby-Roberts}
\affiliation{Department of Physics \& Astronomy, University of Tampa}
\email{jlibbyroberts@ut.edu}

\author[0000-0001-8720-5612]{Joe P.\ Ninan}
\affiliation{\TIFR}
\email{indiajoe@gmail.com}

\author[0000-0001-8127-5775]{Arpita Roy}
\affiliation{Astrophysics \& Space Institute, Schmidt Sciences, New York, NY 10011, USA}
\email{arpita308@gmail.com}

\author[0000-0002-4046-987X]{Christian Schwab}
\affiliation{\Macquarie}
\email{mail.chris.schwab@gmail.com}

\author[0000-0001-7409-5688]{Gudmundur Stefansson}
\affiliation{\UAm}
\email{g.k.stefansson@uva.nl}

\newcommand{\PSUAA}{Department of Astronomy \& Astrophysics, 525 Davey Laboratory, 251 Pollock Road, Penn State, University Park, PA, 16802, USA}
\newcommand{\PSUCEHW}{Center for Exoplanets and Habitable Worlds, 525 Davey Laboratory, 251 Pollock Road, Penn State, University Park, PA, 16802, USA}
\newcommand{\PSUARC}{Astrobiology Research Center, 525 Davey Laboratory, 251 Pollock Road, Penn State, University Park, PA, 16802, USA}
\newcommand{\PSETI}{Penn State Extraterrestrial Intelligence Center, 525 Davey Laboratory, 251 Pollock Road, Penn State, University Park, PA, 16802, USA}
\newcommand{\UA}{Steward Observatory, University of Arizona, 933 N.\ Cherry Ave, Tucson, AZ 85721, USA}
\newcommand{\UAA}{Department of Astronomy and Steward Observatory, University of Arizona, Tucson, AZ 85721, USA}
\newcommand{\Penn}{Department of Physics and Astronomy, University of Pennsylvania, 209 S 33rd St, Philadelphia, PA 19104, USA}
\newcommand{\Caltech}{Department of Astronomy, California Institute of Technology, 1200 E California Blvd, Pasadena, CA 91125, USA}
\newcommand{\STScI}{Space Telescope Science Institute, 3700 San Martin Dr, Baltimore, MD 21218, USA}
\newcommand{\JHU}{Department of Physics and Astronomy, Johns Hopkins University, 3400 N Charles St, Baltimore, MD 21218, USA}
\newcommand{\GoddardESAL}{Exoplanets and Stellar Astrophysics Laboratory, NASA Goddard Space Flight Center, Greenbelt, MD 20771, USA}
\newcommand{\GoddardISTD}{Instrument Systems and Technology Division, NASA Goddard Space Flight Center, Greenbelt, MD 20771, USA}
\newcommand{\GSFC}{NASA Goddard Space Flight Center, Greenbelt, MD 20771, USA}
\newcommand{\NOAO}{U.S. National Science Foundation National Optical-Infrared Astronomy Research Laboratory, 950 N.\ Cherry Ave., Tucson, AZ 85719, USA}
\newcommand{\UW}{Wisconsin address goes here}
\newcommand{\Macquarie}{School of Mathematical and Physical Sciences, Macquarie University, Balaclava Road, North Ryde, NSW 2109, Australia}
\newcommand{\MacquarieCentre}{Astrophysics and Space Technologies Research Centre, Macquarie University, Balaclava Road, North Ryde, NSW 2109, Australia}
\newcommand{\NIST}{National Institute of Standards \& Technology, 325 Broadway, Boulder, CO 80305, USA}
\newcommand{\CUBoulderEE}{Electrical, Computer \& Energy Engineering, 440 UCB, University of Colorado, Boulder, CO 80309, USA}
\newcommand{\CUBoulder}{Department of Physics, 390 UCB, University of Colorado, Boulder, CO 80309, USA}
\newcommand{\JPL}{Jet Propulsion Laboratory, California Institute of Technology, 4800 Oak Grove Drive, Pasadena, California 91109}
\newcommand{\MIT}{Kavli Institute for Astrophysics and Space Research, Massachusetts Institute of Technology, Cambridge, MA, USA}
\newcommand{\UCI}{Department of Physics \& Astronomy, The University of California, Irvine, Irvine, CA 92697, USA}
\newcommand{\Carleton}{Carleton College, One North College St., Northfield, MN 55057, USA}
\newcommand{\Carnegie}{Earth and Planets Laboratory, Carnegie Science, 5241 Broad Branch Road, NW, Washington, DC 20015, USA}
\newcommand{\PSUICS}{Institute for Computational and Data Sciences, Penn State, University Park, PA, 16802, USA}
\newcommand{\PSUCASt}{Center for Astrostatistics, 525 Davey Laboratory, 251 Pollock Road, Penn State, University Park, PA, 16802, USA}
\newcommand{\NESSF}{NASA Earth and Space Science Fellow}
\newcommand{\Princeton}{Department of Astrophysical Sciences, Princeton University, 4 Ivy Lane, Princeton, NJ 08540, USA}
\newcommand{\RUSSELL}{Henry Norris Russell Fellow}
\newcommand{\IAS}{Institute for Advance Study, 1 Einstein Drive, Princeton, NJ 08540, USA}
\newcommand{\Tsinghua}{Department of Astronomy, Tsinghua University, Beijing 100084, China}
\newcommand{\FlatironCCA}{Center for Computational Astrophysics, Flatiron Institute, 162 Fifth Avenue, New York, NY 10010, USA}
\newcommand{\ETH}{ETH Zurich, Institute for Particle Physics \& Astrophysics, Zurich, Switzerland}
\newcommand{\TIFR}{Department of Astronomy and Astrophysics, Tata Institute of Fundamental Research, Homi Bhabha Road, Colaba, Mumbai 400005, India}
\newcommand{\UCLA}{Department of Physics \& Astronomy, University of California Los Angeles, Los Angeles, CA 90095, USA}
\newcommand{\UAm}{Anton Pannekoek Institute for Astronomy, 904 Science Park, University of Amsterdam, Amsterdam, 1098 XH}
\newcommand{\UMd}{Department of Physics \& Astronomy, 1023 University Drive, University of Minnesota Duluth, Duluth, MN 55812, USA}
\newcommand{\UIUC}{Department of Astronomy, University of Illinois at Urbana-Champaign, Urbana, IL 61801, USA}
\newcommand{\Amherst}{Department of Physics and Astronomy, Amherst College, 25 East Drive, Amherst, MA 01002, USA}
\newcommand{\UTAustin}{Department of Astronomy, The University of Texas at Austin, 2515 Speedway, Austin, TX 78712, USA}



\begin{abstract}

We confirm the planetary nature of (1) TOI-5916 b and (2) TOI-6158 b, two Exoplanets Transiting M-dwarf Stars (GEMS), both discovered by the Transiting Exoplanet Survey Satellite (TESS). Both systems were confirmed with ground-based photometry (Red Buttes Observatory and Swope, respectively) and radial velocity data from the Habitable-zone Planet Finder. Their radii are $R_{1}=11.8^{+0.52}_{-0.51}\text{ }R_{\oplus}$ and $R_{2}=10.4^{+2.70}_{-1.11}\text{ }R_{\oplus}$ and masses are $M_{1}=219\pm28\text{ }M_{\oplus}$ and $M_{2}=135^{+19}_{-18}\text{ }M_{\oplus}$. Both planets have Saturn-like densities ($\rho_{1} = 0.73^{+0.14}_{-0.13}\,\text{g\,cm}^{-3}$, $\rho_{2} = 0.66^{+0.41}_{-0.23}\,\text{g\,cm}^{-3}$), which appears to be a growing trend among GEMS systems and, more generally, warm Jupiters. In confirming both of these exoplanets, we add to the growing evidence for a population of Saturn-density planets among the GEMS systems. We also find evidence for a preliminary trend in which GEMS exhibit systematically closer orbits compared to FGK giants. 

\end{abstract}
\keywords{Exoplanet systems (484); Extrasolar gaseous giant planets (509); M dwarf stars (982)}

\section{Introduction} \label{sec:intro}
Giant exoplanets around M-dwarfs (GEMS) have long challenged core accretion planet formation models \citep{Laughlin, Ida05, Burn_2021}, as their unusually high planet-to-star mass ratios cast doubt on whether low-mass protoplanetary disks can accumulate sufficient material within typical lifetimes \citep[$\sim$10 Myr;][]{Pfalzner2024}. An alternative formation pathway---gravitational instability (GI) \citep[GI;][]{Boss06} occurring in the class 0/I disks---may alleviate some of these limitations, although both theories require inward migration to explain the observed short orbital periods of many GEMS. Since the launch of the Transiting Exoplanet Survey Satellite \citep[TESS;][]{Ricker15}, more than 30 transiting GEMS have been discovered, enabling a more detailed discussion of the relationships between their parameters and providing insight into their potentially diverse formation pathways.


GEMS are defined as giant planets orbiting M-dwarfs ($2600 \text{K}- 4000 \text{K}$) having radii of $8~\earthradius \lesssim R_p \lesssim 15~\earthradius$ for transiting planets or $M_p\sin i > 80~\earthmass$ for radial velocity (RV) only detections. The parameter space is defined to select objects that are likely to require runaway gaseous accretion to form \citep{motive}. They are theoretically predicted to be rare because of the difficulties in formation, which is supported by many occurrence rate studies. Previous studies measured occurrence rates as low as $0.194 \pm 0.072\%$\ \citep{Bryant_2023} and as high as $0.27 \pm 0.09\%$ around early-type ($0.45 \ \solmass \leq M_* \leq 0.65 \ \solmass$) M-dwarfs \citep{Gan_23}. A recent study based on a larger sample confirmed the rarity of GEMS, reporting occurrence rates of $0.067\%\pm0.047\%$ for early-type M dwarfs, $0.139\%\pm0.069\%$ for mid-type M dwarfs, and $0.032\%\pm0.032\%$ for late-type M dwarfs (Glusman et al. 2025, submitted).

Preliminary trends in the GEMS sample suggest diverse formation pathways that may sculpt the observed properties of this sample. Most GEMS host stars exhibit either solar or super-solar metallicities, the latter being consistent with core accretion being favored in metal-rich protoplanetary disks \citep{Kagetani2023, Han2024, Gan2025}, in line with trends observed for hot Jupiters \citep[see][and references therein]{Osborn2020}. In contrast, the discovery of several super-Jupiters orbiting M dwarfs points to GI as a more likely formation mechanism in those environments \citep{Boss2023,Hotnisky2025}. Atmospheric characterization of GEMS has been enabled by the James Webb Space Telescope \citep[JWST;][]{JWST}, using transmission spectroscopy.  Additionally, some of the super-Jupiters around M-dwarfs exhibit eccentric orbits \citep{Bryant2024, Hotnisky2025}, hinting at dynamically hot migration processes such as planet–planet scattering \citep{Rasio1996, Dong2025}. Given the diversity of apparent formation scenarios within the currently small GEMS sample, expanding the population is crucial to draw more robust conclusions. 

Since the launch of the \textit{Searching for GEMS Survey} \citep{motive}, we have discovered and confirmed dozens of new GEMS, increasing the number of known transiting GEMS to 35, which includes the two GEMS systems confirmed in this work. This expanded sample enables more reliable studies of property trends and occurrence rates, which is essential to constrain planet formation pathways. In this paper, we confirm the planetary nature of TOI-5916 b and TOI-6158 b, both initially identified as planet candidates by TESS. They represent valuable additions to the growing sample of transiting GEMS. We confirmed TOI-5916 b with ground-based photometry from the Red Buttes Observatory (RBO), as well as with radial velocity (RV) measurements from the Habitable-zone Planet Finder (HPF). Additionally, we confirmed TOI-6158 b, a planet with a grazing transit, with ground-based photometry from the Henrietta Swope telescope at Las Campanas Observatory, as well as with RV measurements from HPF. The details of these observations are given in Section \ref{sec:observations} of this paper. In Section \ref{sec: ruling out}, we discuss how we rule out nearby bright stellar companions in the observed systems. Stellar parameters for each system are derived in Section \ref{sec: stellar params}. In Section \ref{sec: joint fit} we discuss the joint fitting of the RV and photometry data, as well as the tools that were used to perform this joint fit. Section \ref{sec: discussion} discusses how these systems fit into the mosaic of GEMS systems and what it shows, and Section \ref{sec: conclusion} summarizes our findings in this analysis.

\section{Observations} \label{sec:observations}

\subsection{TESS Photometry \& Anomalous Background Transit} \label{subsec: TESS} 
\begin{figure*}
    \centering
    \includegraphics[width=\textwidth]{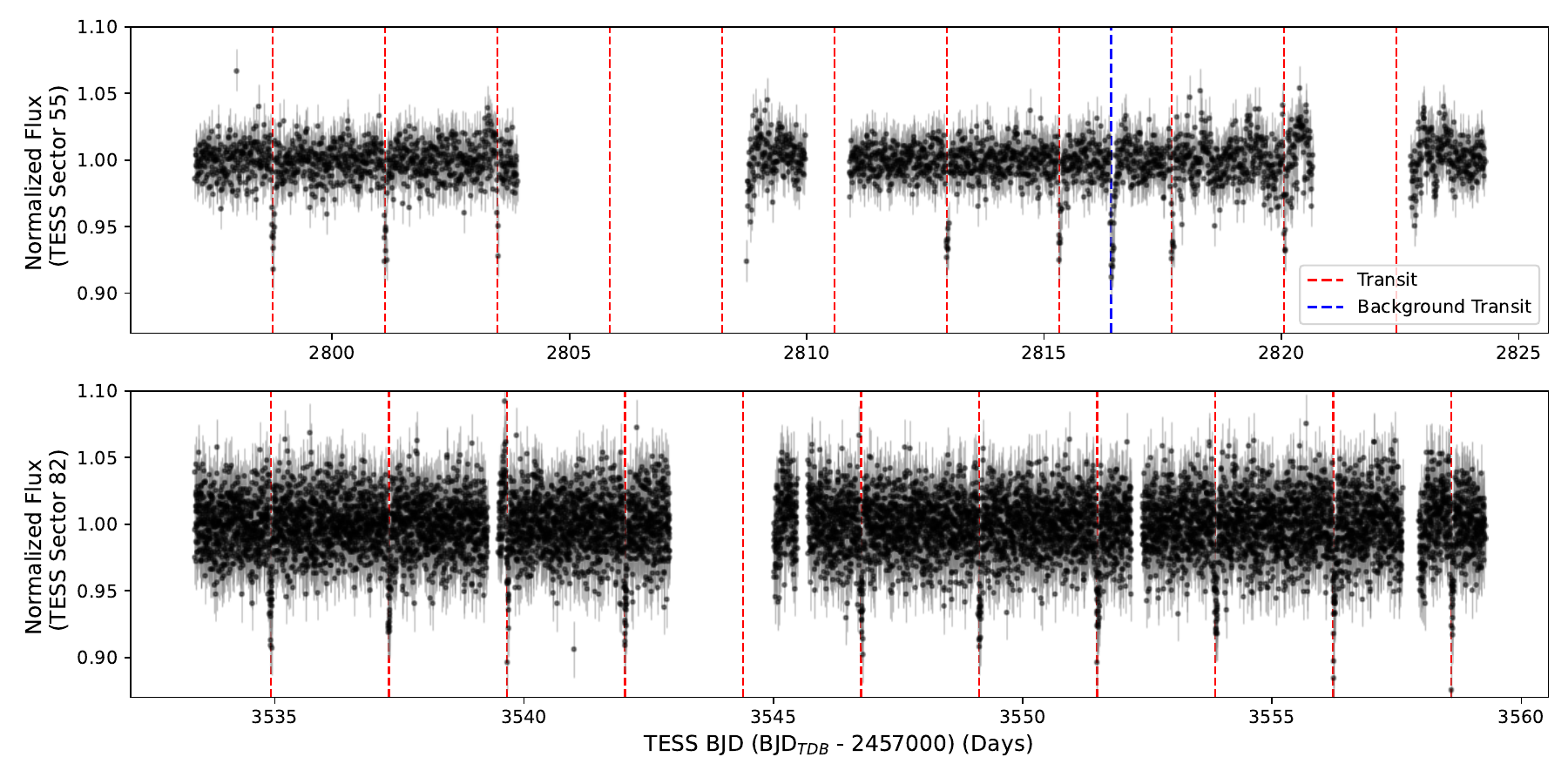}
    \caption{TGLC calibrated aperture light curve of TOI-5916 for TESS Sectors 55 (600s cadence) and 82 (200s cadence). Red lines correspond to the observed transits of TOI-5916 b. The blue line corresponds to an anomalous background transit that was observed by TESS during its observation of TOI-5916 (see Section \ref{sec: joint fit} for more on this background transit).}
    \label{fig:5916_Photometry}
\end{figure*}

\begin{figure*}
    \centering
    \includegraphics[width=\textwidth]{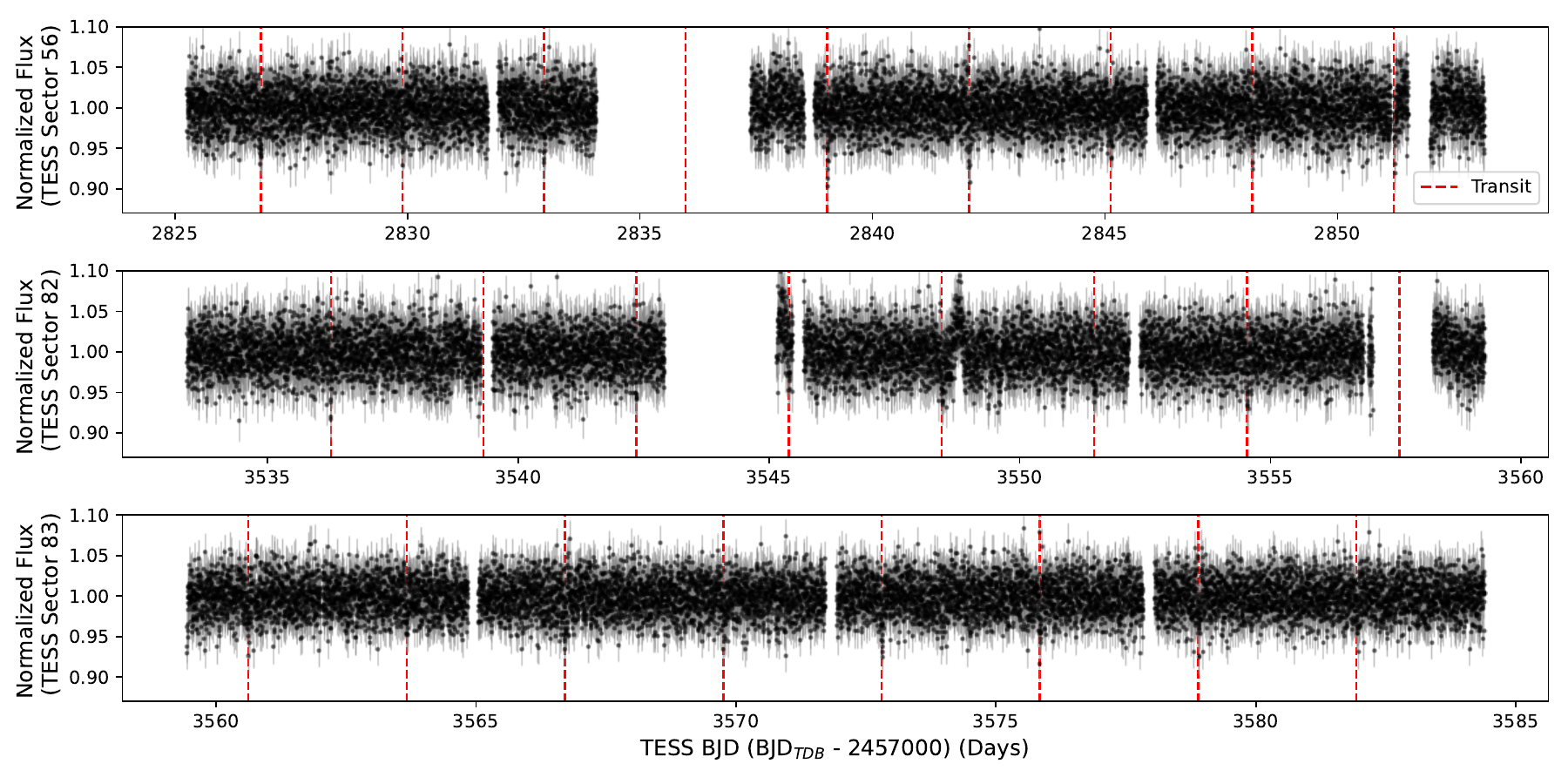}
    \caption{Similiar to Figure \ref{fig:5916_Photometry}, but for TOI-6158.}
    \label{fig:6158_Photometry}
\end{figure*}

TOI-5916 b and TOI-6158 b were first identified as planet candidates via the TESS Faint Star Search \citep{faint-star}, which was based on the Quick Look Pipeline \citep[QLP;][]{QLP,Kunimoto22}. TOI-5916 (\textit{TIC 305506996}) was observed from 2022 August 5 to 2022 September 1 in TESS sector 55 and from 2024 August 10 to 2024 September 5 in TESS sector 82 (Figure \ref{fig:5916_Photometry}). Data from TESS sector 55 were collected using Camera 1 with an exposure of 600~s. Data from TESS sector 82 were collected using Camera 1 with an exposure time of 200~s. TOI-6158 (\textit{TIC 404456775}) was observed in sectors 56, 82, and 83 (Figure \ref{fig:6158_Photometry}). Sector 56 was observed from 2022 September 2 to 2022 September 30 with an exposure time of 600~s. Sector 82 was observed from 2024 August 10 to 2024 September 5 with an exposure time of 200~s. Sector 83 was observed on 2024 September 5 to 2024 September 30 with an exposure time of 200~s.

\clearpage
\begin{figure*}
    \centering
    \includegraphics[width=\textwidth]{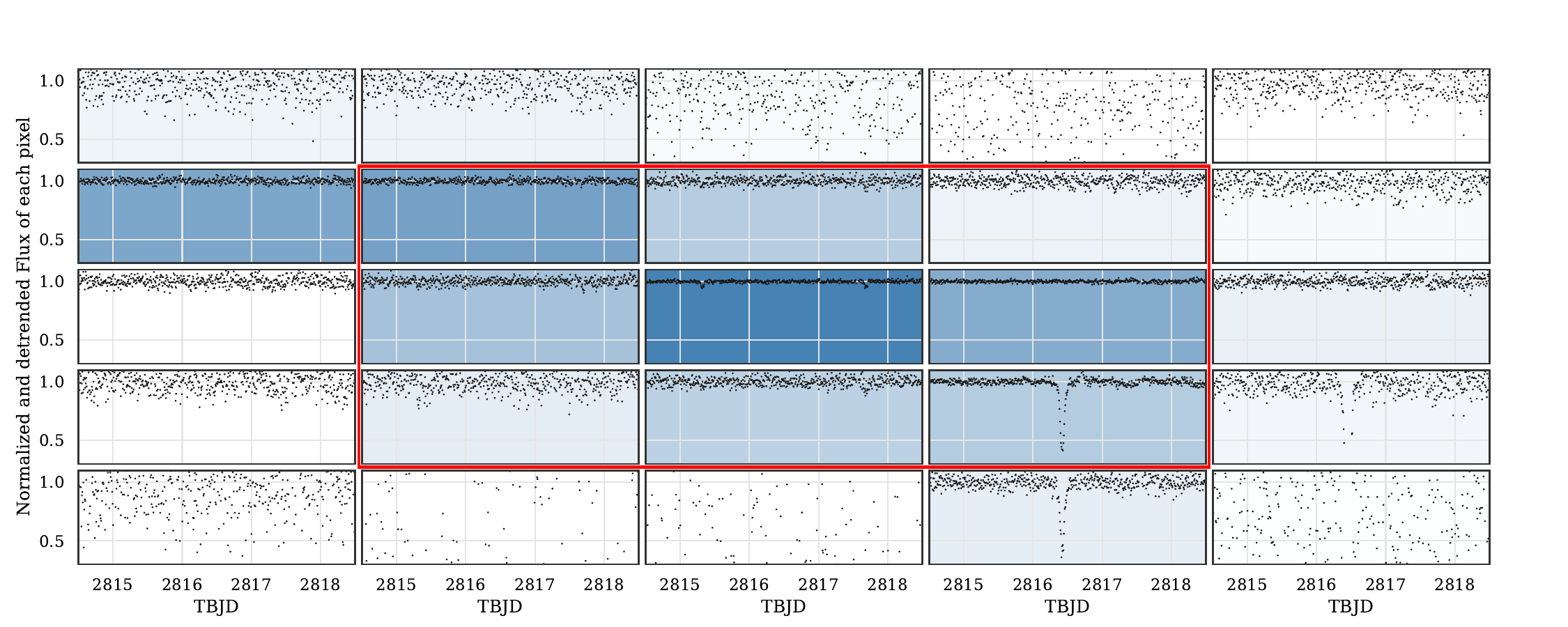}
    \caption{Pixel-by-pixel analysis of the anomalous transit observed in TGLC Sector 55 data. The red box indicates the $3\times3$ aperture used by TGLC, while the blue shading represents the relative baseline fluxes of individual pixels after subtracting background stars. Each panel displays the normalized and detrended light curves obtained by removing flux contributions from neighboring stars. A genuine on-target signal would produce consistent transit depths across all pixels, as observed for TOI-5916 b. Therefore, the transit anomaly detected only in a few pixels is unlikely to originate from TOI-5916 itself.}
    \label{fig:5916_Sec55Pixels}
\end{figure*}




We extracted the TESS-Gaia Light Curve \citep[TGLC;][]{Han_2023} from the TESS Full-Frame Images (FFIs), employing point-spread function modeling to remove contamination from nearby stars. This approach effectively corrects for the dilution caused by light from neighboring stars and background sources, and its accuracy has been statistically validated through comparisons with high-resolution ground-based observations and Kepler data \citep{Han2025}. We ran the $\texttt{tglc}$ python package on FFI cuts of $90 \times 90$ pixels, and extracted the calibrated aperture light curve ($\texttt{cal\_aper\_flux}$). We excluded data points marked by TESS data quality flags or TGLC data quality flags in our analysis.

In the case of the TOI-5916 b Sector 55 TESS data, we note that an anomalous background transit was detected in these data, as shown by the blue line in Figure \ref{fig:5916_Photometry}. To further investigate this anomalous transit, a pixel-by-pixel analysis was conducted on these collected data, which is shown in Figure \ref{fig:5916_Sec55Pixels}. This pixel-level analysis shows that the observed dip in brightness was very unlikely to be from the target star, TOI-5916. In addition to this, there are also no Gaia DR3 objects near the center of the anomaly transit, precluding it from being a nearby star. We thus excise a $\pm$ 0.2 day region around the point of deepest transit in order to ensure that this background transit does not impact the results of our joint fit.


\subsection{RBO Photometry} \label{subsec: RBO}

\begin{figure*}
    \centering
    \includegraphics[width=\textwidth]{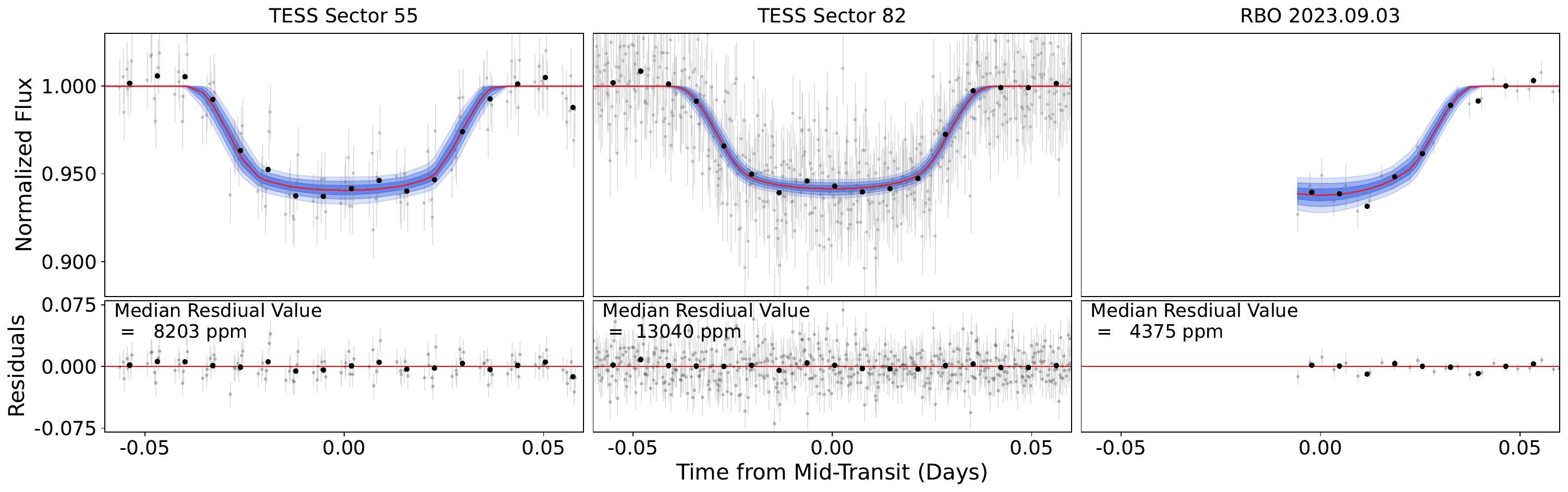}
    \caption{The phase-folded light curve for TOI-5916. The gray data points are the raw photometry data. The black data points are the 10-minute binned data points. The red line represents the median (best joint fit model). The confidence intervals, ranging from 1$\sigma$ to 3$\sigma$, are shown in decreasing intensity of blue. For TESS Sector 55, the exposure time was 600~s, where as for TESS Sector 82 it was 200~s. The exposure time for the RBO observation was 240~s.}
    \label{fig:5916_PFLC}
\end{figure*}

We observed a partial transit of TOI-5916 with the telescope at Red Buttes Observatory \citep[RBO;][]{Kasper16} in Wyoming, USA. The observation started mid-transit due to the target reaching the airmass limit while rising. The telescope is a 0.6~m Ritchey-Chrétien Cassegrain telescope of f/8.43. The observation of the partial transit was conducted on 2023 September 3, with an exposure time of 240~s. During this observation, data were taken with the Alta F16 camera, which used the Bessell $I$ filter, at an airmass of $2.75$ – $1.20$.  Our images showed a typical seeing-limited full width at half maximum (FWHM) in the point-spread function (PSF) of about $2^{\prime \prime}$. The raw photometry was reduced and the light curve was extracted using the \texttt{Python} pipeline described in \S2.1.1 of \cite{KanodiaSixGEMS}. The phase-folded light curves of these data, together with these data from TESS, are shown in Figure~\ref{fig:5916_PFLC}.



\subsection{Swope Photometry} \label{subsec: SWOPE}

\begin{figure*}
    \centering
    \includegraphics[width=\textwidth]{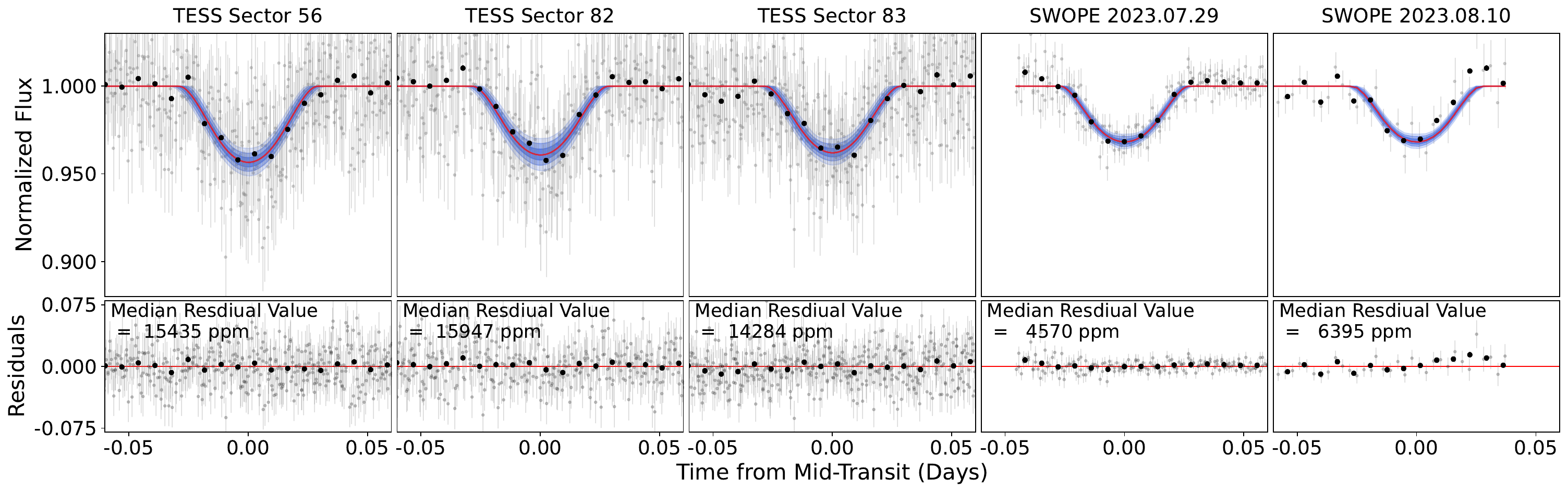}
    \caption{Same as Figure \ref{fig:5916_PFLC}, but for TOI-6158. The exposure time for TESS Sectors 56, 82, and 83 were 600~s, 200~s, and 200~s respectively. The exposure time for the Swope observations were 40~s and 240~s respectively.}
    \label{fig:6158_PFLC}
\end{figure*}
We observed TOI-6158 with the Henrietta Swope Telescope, a 1.0\,m reflector, at Las Campanas Observatory in Chile. The telescope is an f/7 Ritchey-Chrétien designed to achieve a flat field and a zero Petzval sum. 

On the night of 2023 July 30, we observed a transit of TOI-6158b in the SDSS $i$ band over a total of 7 hours.  During this time, the target rose from an airmass of 2.60, peaked at 1.42, and increased to 2.76.  We defocused the telescope to yield PSF FWHM of roughly $4^{\prime \prime}$, and used an exposure time of 40s.

On the night of 2023 August 10, we observed a transit of TOI-6158 b across a total of 4.1 hours in the SDSS $g$ band.  The star was at an airmass of 1.43 at the start of observations, and was at airmass 2.78 at the conclusion.  The telescope was again defocused to a PSF FWHM$\sim 4^{\prime \prime}$, but we used 240s exposures for the $g$-band lightcurve.

Data reduction and standard aperture photometry was performed using \texttt{AstroImageJ} \citep[AIJ;][]{KACollins}. The light curves of the Swope data, together with the phase-folded data from TESS, are shown in Figure~\ref{fig:6158_PFLC}.

\subsection{Habitable-zone Planet Finder Spectroscopy} \label{subsec: HPF}

\begin{deluxetable}{cccc}
\tablecaption{TOI-5916 HPF RVs \label{tab:RV}}
\tablehead{
\colhead{BJD} & \colhead{RV (m/s)} & \colhead{$\sigma$ (m/s)}
}
\startdata
2460146.94877 & -233 & 45\\
2460162.90043 & 18 & 50 \\
2460187.83757 & -154 & 47 \\
2460188.83400 & 64 & 36 \\
2460189.83157 & -202 & 44 \\
2460190.83147 & 67 & 33 \\
2460192.65831 & -243 & 57 \\
2460196.64664 & -207 & 41 \\
2460202.63831 & 44 & 51 \\
2460212.60655 & 29 & 40 \\
2460216.75225 & -37 & 48 \\
2460222.58148 & -177 & 39 \\
\enddata
\tablecomments{Median SNR: 25}
\end{deluxetable}

\begin{deluxetable}{ccc}
\tablecaption{TOI-6158 HPF RVs \label{tab:RV_6158}}
\tablehead{
\colhead{BJD} & \colhead{RV (m/s)} & \colhead{$\sigma$ (m/s)}
}
\startdata
2460133.82714 & 96 & 42 \\
2460142.80933 & 53 & 35  \\
2460146.79170 & -37 & 33  \\
2460149.78594 & 24 & 38 \\
2460155.77470 & -99 & 28 \\
2460161.75108 & -61 & 37 \\
2460166.93508 & 79 & 32 \\
2460171.72259 & -188 & 42 \\
2460192.86364 & -85 & 33 \\
2460193.66318 & 0 & 34 \\
2460193.85904 & -33 & 39 \\
2460195.86019 & -161 & 33 \\
2460215.79442 & 49 & 45 \\
2460215.79442 & -112 & 92 \\
2460303.55709 & 33 & 31 \\
2460455.95661 & -30 & 38 \\
2460456.93657 & -40 & 44 \\
2460457.94410 & -79 & 31 \\
2460460.93369 & -55 & 39 \\
2460461.92148 & 79 & 48 \\
2460468.91016 & 59 & 30 \\
2460478.87784 & -192 & 42 \\
2460479.87819 & -101 & 45 \\
2460480.88057 & 24 & 39 \\
2460484.87602 & -108 & 50 \\
2460486.86828 & 84 & 45 \\
2460488.84826 & -37 & 35 \\
2460501.82024 & 150 & 47 \\
\enddata
\tablecomments{Median SNR: 24}
\end{deluxetable}

We observed TOI-5916 and TOI-6158 with the Habitable-zone Planet Finder Spectrometer \citep[HPF;][]{Mahadevan12, Mahadevan14, Stefansson2016, Kanodia18}. HPF is a high-resolution near-infrared spectrograph (8080–12,780 \AA) mounted on the 10.0-m Hobby-Eberly Telescope \citep[HET;][]{Ramsey98, Hill2021} at McDonald Observatory in Texas, USA. The raw data were corrected using the \texttt{HxRGproc} routines \citep{HxRGproc}, and the procedure outlined in \citet{Stefansson2020} was used to obtain the wavelength solutions. Radial velocities were extracted through use of a custom version of the \texttt{SERVAL} pipeline \citep{Zechmeister18, Stefansson2023}, and all spectra were corrected to the barycentric frame via \texttt{barycorrpy} \citep{Wright14, Kanodia18b}.

\begin{figure*}
    \centering
    \includegraphics[width=\textwidth]{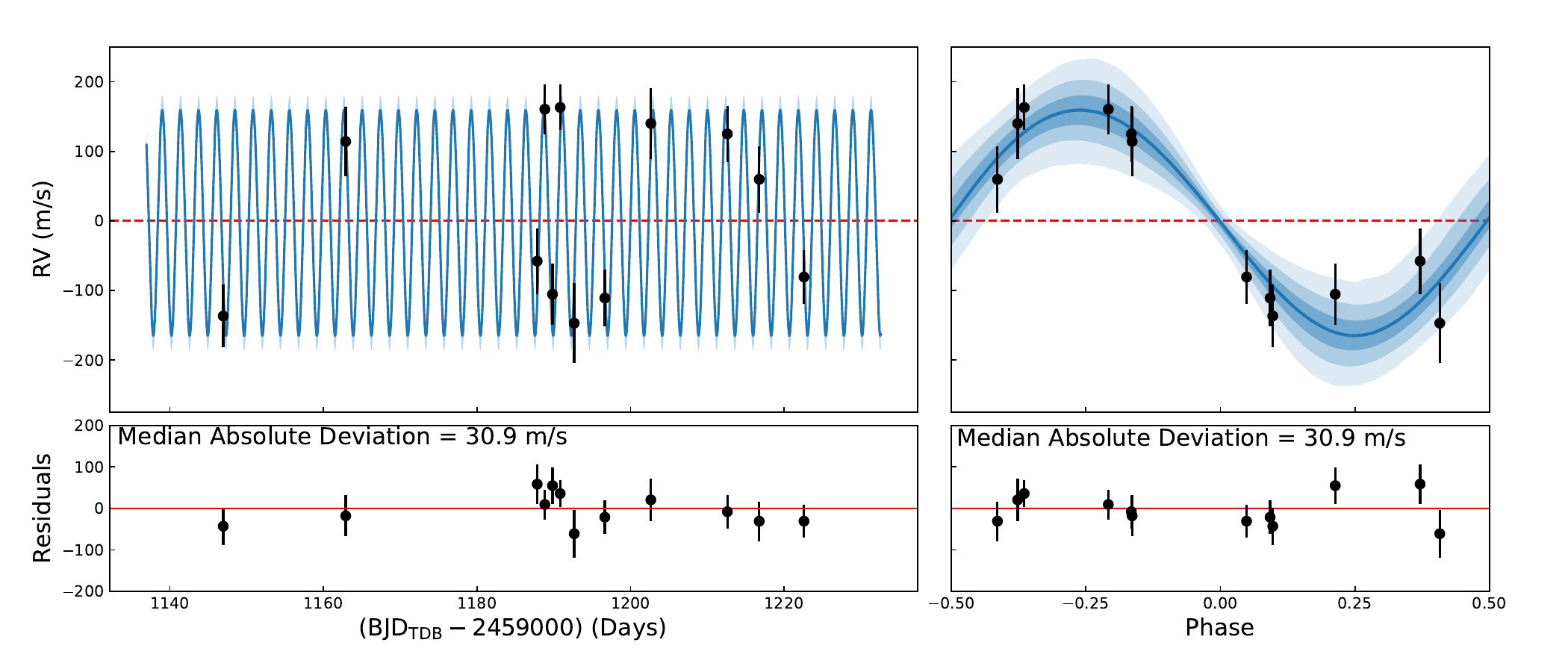}
    \caption{Binned HPF RV data for TOI-5916. \textbf{Left:} The time series of the RV data binned by each night (black points), with the median model value in solid blue and the light blue region being the $1\sigma$ region. \textbf{Right:} The phase-folded RV curve for the binned data. The median model value is shown in solid blue and the $1\sigma$, $2\sigma$, and $3\sigma$ confidence intervals are shown in descending intensities of blue.}
    \label{fig:5916_Combined_RV}
\end{figure*}

For TOI-5916, binned RV data were used, as shown in Table \ref{tab:RV}. Each binned RV value represents the weighted average of two consecutive exposures, each of which has an exposure time of 945~s. The median signal-to-noise ratio (SNR) for these data is 24.84. The time series plot of the data, as well as the phase-folded plot, are shown in Figure \ref{fig:5916_Combined_RV}.

\begin{figure*}
    \centering
    \includegraphics[width=\textwidth]{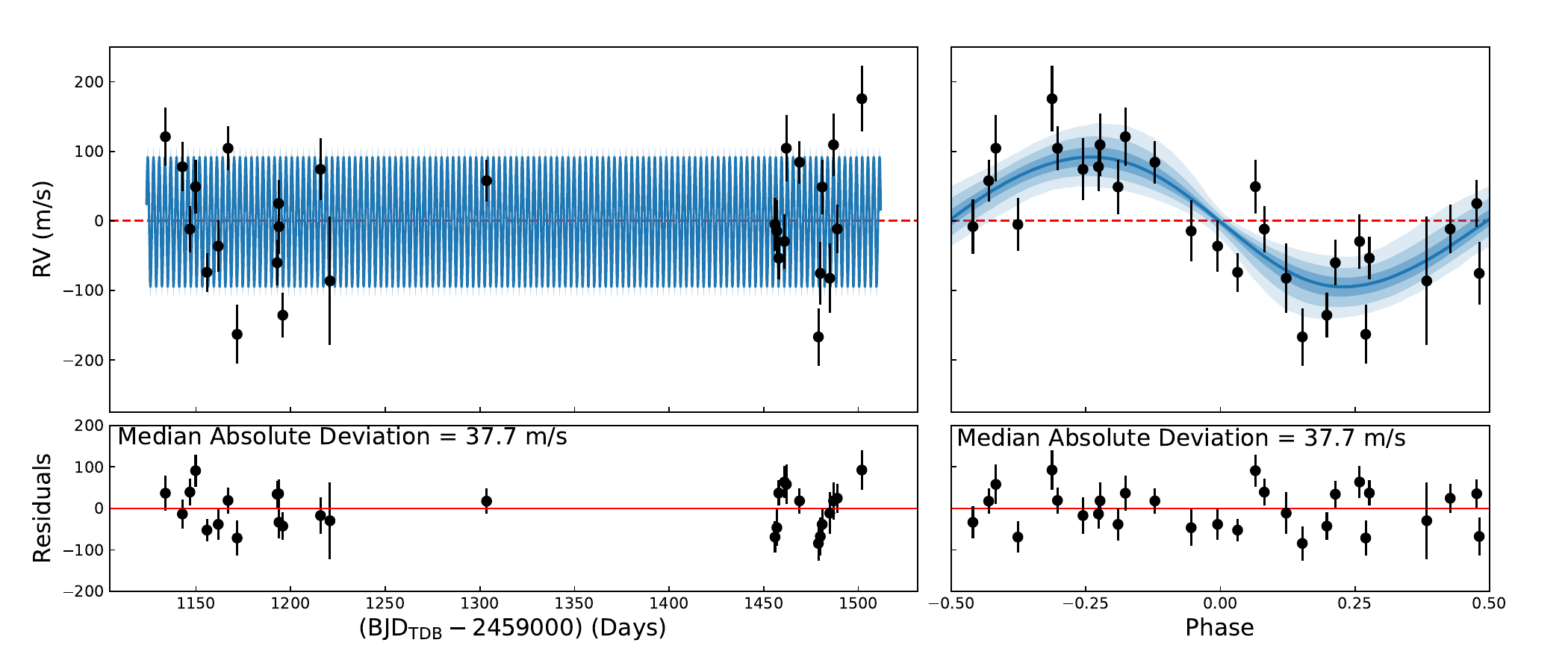}
    \caption{Same as Figure \ref{fig:5916_Combined_RV}, but for TOI-6158.}
    \label{fig:6158_Combined_RV}
\end{figure*}

For TOI-6158, binned data were also used. We have 56 exposures taken over 28 nights, with two exposures taken each night. Each exposure was 969~s and has a median per-pixel SNR of 23.63 at 10000 \AA{}. The selected exposures are listed in Table \ref{tab:RV_6158}. Similar to the case with TOI-5916, the time series plot of the data, as well as the phase-folded plot, are shown in Figure \ref{fig:6158_Combined_RV}.


\section{Ruling out nearby Stellar Companions} \label{sec: ruling out}
\subsection{ShaneAO}
We observed both stars using the ShARCS camera \citep{srinath_swimming_2014} on the Shane 3-meter telescope at Lick Observatory in California, USA, operated by the University of California. TOI-5916 and TOI-6158 were observed on the nights of 2024 May 25 and 2024 May 26, respectively. Observations were conducted in Laser Guide Star (LGS) mode without dithering due to instrument failures. To estimate the sky background, we adopted an on-target and off-target observing strategy. TOI-5916 was observed on-target for a total of 600\,s and off-target for 400\,s in the $K_s$ filter. TOI-6158 was observed on-target for 1050\,s and off-target for 300\,s, also in the $K_s$ filter. These data were reduced using our own customized pipeline \citep[see details in \S2.5 of][]{Canas2020}. We detected no background sources with separations $>0.5''$ and $\Delta K_s < 3.5$ for TOI-5916, and none with separations $>1.0''$ and $\Delta K_s < 2.5$ for TOI-6158.

\subsection{NESSI}

\begin{figure*}
    \centering
    \includegraphics[width=0.5\textwidth]{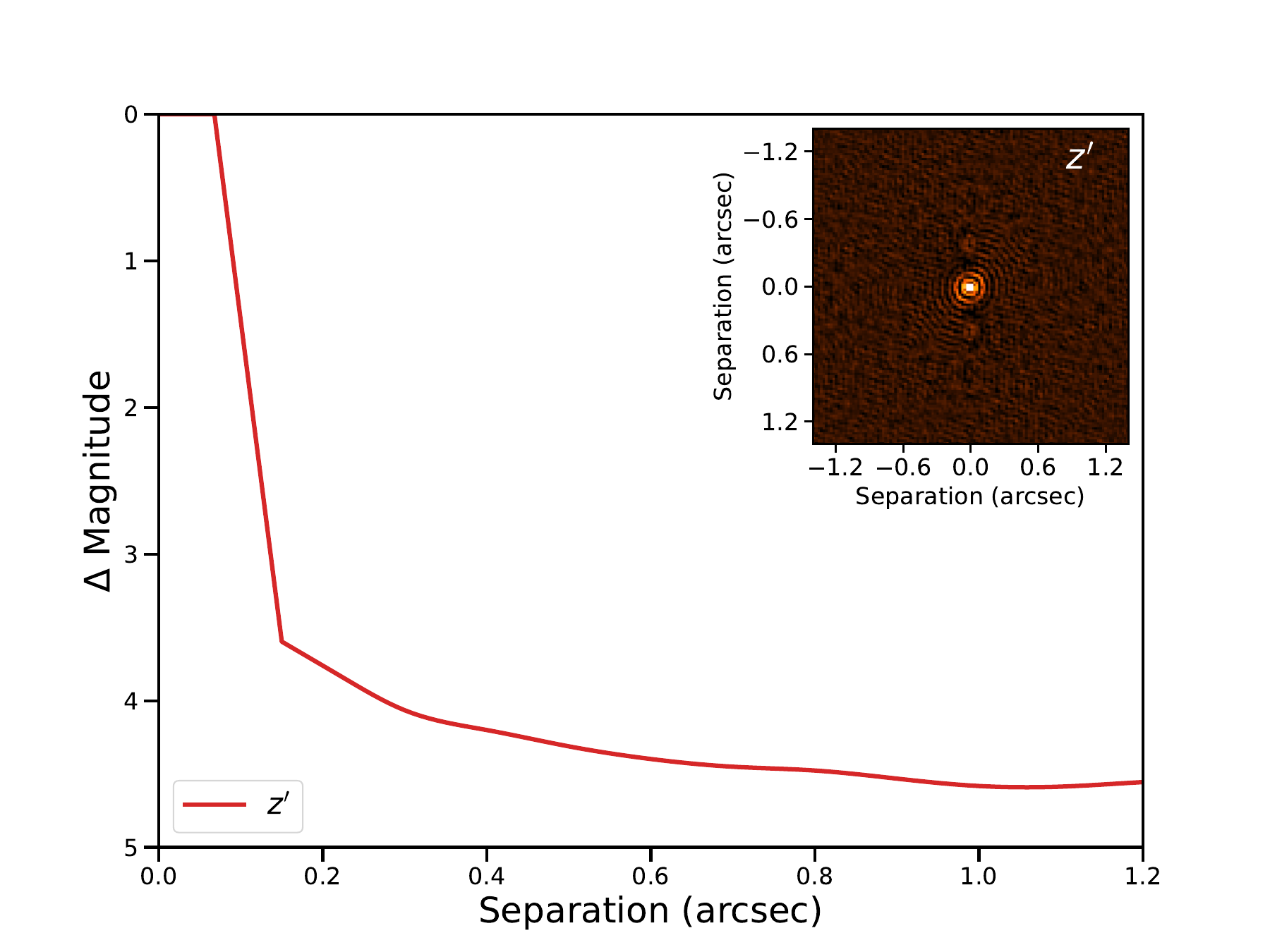}\includegraphics[width=0.5\textwidth]{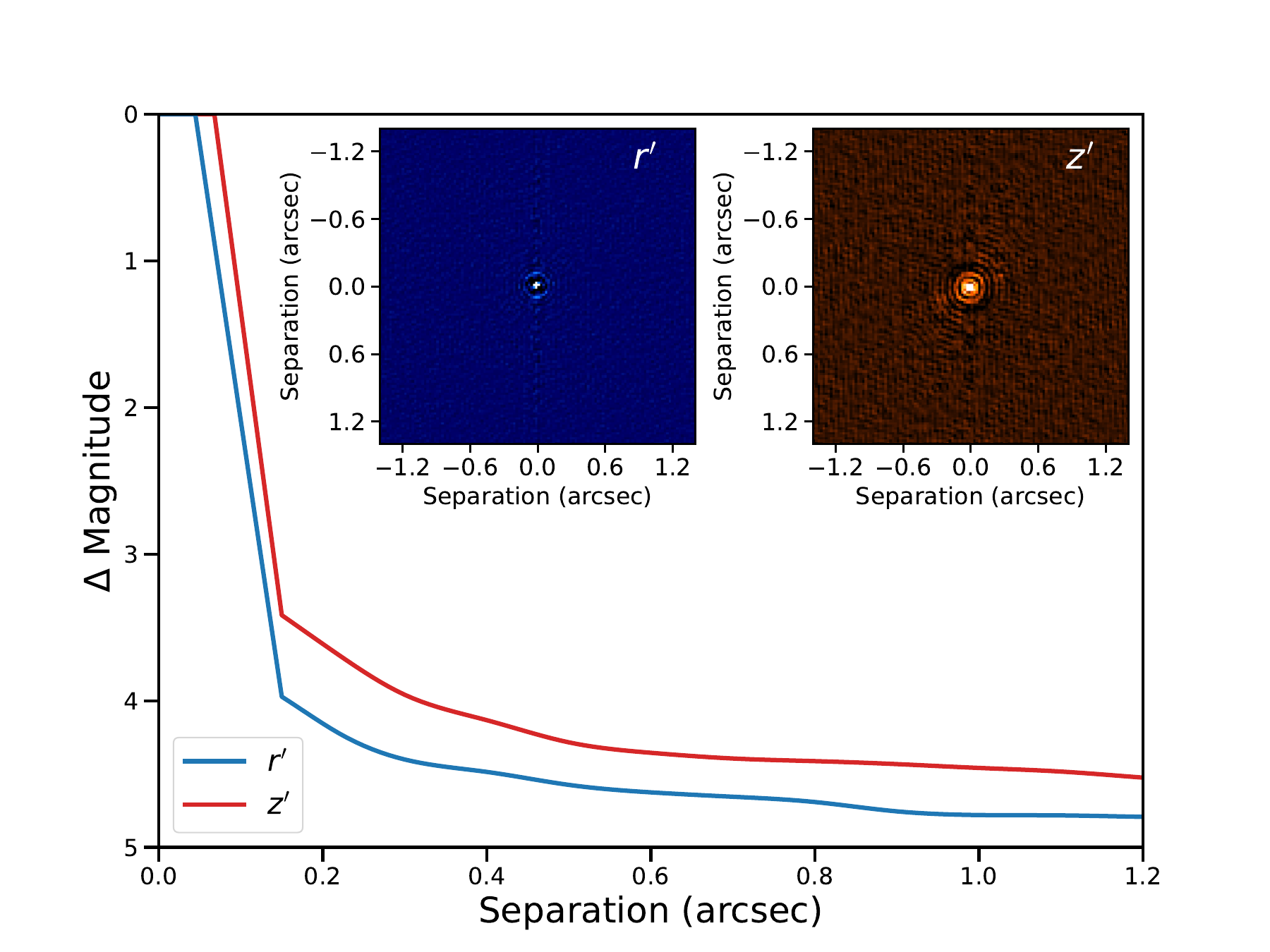}
    \caption{NESSI speckle images and contrast limits for TOI-5916 (left) and TOI-6158 (right). The 5-$\sigma$ contrast limits are shown in blue and red for the r' and z' filters, respectively, and cutouts of the reconstructed speckle images are shown as insets in the upper corner of each plot. No nearby companions or background sources are detected for either star.}
    \label{fig:NESSI}
\end{figure*}

Both TOI-5916 and TOI-6158 were also observed at high spatial resolution with the NN-Explore Exoplanet Stellar Speckle Imager \citep[NESSI;][]{Scott2018} on the 3.5\,m WIYN\footnote{The WIYN Observatory is a joint facility of the NSF's National Optical-Infrared Astronomy Research Laboratory, Indiana University, the University of Wisconsin-Madison, Pennsylvania State University, Purdue University and Princeton University.} Telescope at Kitt Peak National Observatory in Arizona.  Both observations took place on the night of 2023 September 9. For each star, the NESSI observations were carried out as a sequence of 9 image sets each consisting of 1000 40 ms diffraction-limited frames, taken using the SDSS $z^\prime$ and $r^\prime$ filters with the NESSI red and blue cameras, respectively. In \autoref{fig:NESSI}, we show the achieved 5-$\sigma$ contrast curves and the resulting speckle images, which are reconstructed following the methods given in \citet{Howell2011}. The reliability of the $r'$ contrast limits for TOI-5916 is questionable because an artifact was identified in the image reconstruction, so we limit our analysis to the $z'$ data for this target. The speckle data are sufficient to rule out nearby companions and foreground and background sources down to magnitude limits of  $\Delta$m$_{z'}=3.8$ at 0.2'' and $\Delta$m$_{z'}=4.2$ at 1'' for TOI-5916 and $\Delta$m$_{r'}=4.2$ and $\Delta$m$_{z'}=3.6$ at 0.2'' and $\Delta$m$_{r'}=4.8$ and $\Delta$m$_{z'}=4.5$ at 1'' for TOI-6158.

\section{Stellar Parameters} \label{sec: stellar params}

\subsection{HPF-SpecMatch}\label{sec:hpfspecmatch}
We used the \href{https://gummiks.github.io/hpfspecmatch/}{\texttt{HPF-SpecMatch}} package \citep[][]{Stefansson2020} to estimate the stellar effective temperature ($T_{\text{\tiny eff}}$), surface gravity ($\log g_\star$), metallicity ([Fe/H]), and rotational velocity ($v\sin i_\star$) for our host stars. \texttt{HPF-SpecMatch} calculates these spectroscopic parameters using a weighted linear combination of the five best-matching spectra from a library of well-characterized stars \citep[e.g.,][]{Yee2017}. In this work, the library contained 100 stars that spanned $2700\mathrm{K} \le T_{\text{\tiny eff}} \le 4500~\mathrm{K}$, $4.63<\log g_\star < 5.26$, and $-0.49 < \mathrm{[Fe/H]} < 0.53$. We applied \texttt{HPF-SpecMatch} to all spectra using order index 5 ($8534-8645$ \AA{}) because of the minimal telluric contamination in that region. The reported uncertainties are the standard deviation of the residuals of each spectroscopic parameter based on a leave-one-out cross-validation procedure applied to the stellar library \citep[see Section 3.1 in][]{Stefansson2020}. We determined (i) $T_{\text{\tiny eff}}=3541\pm59$ K, $\log g_\star=4.79\pm0.04$, and $\mathrm{[Fe/H]}=0.04\pm0.16$ for TOI-5916 and (ii) $T_{\text{\tiny eff}}=3467\pm59$ K, $\log g_\star=4.81\pm0.04$, and $\mathrm{[Fe/H]}=0.11\pm0.16$ for TOI-6158. \texttt{HPF-SpecMatch} placed an upper limit of $v\sin i_\star<2~\mathrm{km~s^{-1}}$ for both TOI-5916 and TOI-6158. These values are listed in \autoref{tab:stellarparam}.

\begin{deluxetable*}{lcccc}
\tabletypesize{\fontsize{8}{11}\selectfont}
\tablecaption{Summary of Stellar Parameters for TOI-5916 \& TOI-6158. \label{tab:stellarparam}}
\tablehead{
\colhead{~~~Parameter} &  
\colhead{Description} &
\colhead{TOI-5916 Value} &
\colhead{TOI-6158 Value} &
\colhead{Reference} 
}
\startdata
\multicolumn{5}{l}{\hspace{-0.2cm} Main Identifier:}  \\
~~~TOI & TESS Object of Interest & 5916 & 6158 & TESS Mission \\
~~~TIC & TESS Input Catalog & 305506996 & 404456775 & Stassun \\
~~~Gaia DR3 & Gaia Data Release 3 & 1741429438313012608 & 1775027485705816960 & Gaia DR3 \\
\multicolumn{5}{l}{\hspace{-0.2cm} Equatorial Coordinates, Proper Motion, \& Parallax:}  \\
~~~$\alpha$ & Right Ascension (RA) & 21h41m11.88s & 22h15m34.66s & Gaia DR3 \\  
~~~$\delta$ & Declination (Dec) & +09d35m56.59s & +16d17m07.41s & Gaia DR3 \\
~~~$\mu_{\alpha}$ & Proper Motion (RA, mas/yr) & -8.933 $\pm$ 0.120 & 62.883 $\pm$ 0.118 & Gaia DR3 \\
~~~$\mu_{\delta}$ & Proper Motion (Dec, mas/yr) & -25.592 $\pm$ 0.127 & -35.263 $\pm$ 0.117 & Gaia DR3 \\
~~~$\varpi$ & Parallax (mas) & 4.98 $\pm$ 0.08  & 5.50 $\pm$ 0.11 & Gaia DR3 \\
~~~$A_{\nu\text{, max}}$ & Maximum Visual Extinction & 0.030 & 0.040 & Green \\
\multicolumn{5}{l}{\hspace{-0.2cm} Optical \& Near-Infrared Magnitudes:}  \\
\hspace{.3cm}$J$ & 2MASS $J$ mag & 13.118 $\pm$ 0.023 & 13.009 $\pm$ 0.019 & 2MASS \\
\hspace{.3cm}$H$ & 2MASS $H$ mag & 12.016 $\pm$ 0.022 & 12.378 $\pm$ 0.025 & 2MASS \\
\hspace{.3cm}$K_S$ & 2MASS $K_S$ mag & 12.235 $\pm$ 0.029  & 12.15 $\pm$ 0.02 & 2MASS \\
\hspace{.3cm}$g$ & PanSTARRS g mag & 17.268 $\pm$ 0.014 & 17.447 $\pm$ 0.015 & PS1 DR2\\
\hspace{.3cm}$r$ & PanSTARRS r mag & 16.096 $\pm$ 0.009 & 16.270 $\pm$ 0.008 & PS1 DR2\\
\hspace{.3cm}$i$ & PanSTARRS i mag & 14.997 $\pm$ 0.005 & 15.045 $\pm$ 0.008 & PS1 DR2\\
\hspace{.3cm}$z$ & PanSTARRS z mag & 14.510 $\pm$ 0.010 & 14.495 $\pm$ 0.009 & PS1 DR2\\
\hspace{.3cm}$y$ & PanSTARRS y mag & 14.291 $\pm$ 0.014 & 14.229 $\pm$ 0.008 & PS1 DR2\\
\hspace{.3cm}$W1$ & WISE1 mag & 12.091 $\pm$ 0.023 & 12.024 $\pm$ 0.023 & WISE \\
\hspace{.3cm}$W2$ & WISE2 mag & 12.016 $\pm$ 0.022 & 11.93 $\pm$ 0.023 & WISE \\
\hspace{.3cm}$W3$ & WISE3 mag & 11.603 $\pm$ 0.216 & 11.435 $\pm$ 0.267 & WISE \\
\multicolumn{5}{l}{\hspace{-0.2cm} Spectroscopic Parameters:}  \\ 
~~~$T_{\text{\tiny eff}}$ & Effective Temperature (K) & 3541 $\pm$ 59 & 3467 $\pm$ 59 & This work \\
~~~[Fe/H] & Metalicity (dex) & 0.04 $\pm$ 0.16 & 0.11 $\pm$ 0.16 & This work \\
~~~$\log g_\star$ & Surface Gravity (cgs units) & 4.79 $\pm$ 0.04 & 4.81 $\pm$ 0.04 & This work \\
~~~$v\sin i$ & Rotational Velocity (km/s) & $<2~\mathrm{km~s^{-1}}$ & $<2~\mathrm{km~s^{-1}}$ & This work \\
\multicolumn{5}{l}{\hspace{-0.2cm} Model Dependent Stellar SED \& Isochrome fit Parameters:}  \\ 
~~~$M_\star$ & Mass ($\solmass$) & 0.518$\pm$ 0.023 & 0.503$^{+0.022}_{-0.023}$ & This work \\
~~~$R_\star$ & Radius ($\solradius$) & 0.487 $^{+0.013}_{-0.012}$ & 0.476$^{+0.013}_{-0.013}$ & This work \\
~~~$L_\star$ & Luminosity ($\sollum$) & 0.0349 $\pm$ 0.001 & 0.03082$^{+0.001}_{-0.00099}$ & This work \\
~~~$\rho_\star$ & Density (g/cm$^3$) & 6.17 $^{+0.54}_{-0.48}$ & 6.66 $^{+0.49}_{-0.47}$ & This work \\
~~~$A_{\nu}$ & Visual Extinction (mag) & 0.014 $\pm$ 0.01 & 0.018$^{+0.014}_{-0.013}$ & This work \\
~~~d & Distance (pc) & 196 $\pm$ 1.6 & 181.6 $^{+2.2}_{-2.1}$ & This work \\
\enddata
\tablecomments{References: Stassun \citep{TESS_Catalog}, Gaia DR3 \citep{Gaia_2023}, Green \citep{Green2019}, 2MASS \citep{2MASS}, PS1 DR2 \citep{PS1_DR2}, WISE \citep{WISE}}
\end{deluxetable*}

\begin{deluxetable*}{lccccc}
\tabletypesize{\fontsize{8}{11}\selectfont}
\tablecaption{Summary of Parameters for TOI-5916 b \& TOI-6158 b. \label{tab:PlanetParams}}
\tablehead{
\colhead{~~~Parameter}&  
\colhead{Description}&
\colhead{TOI-5916 b}&
\colhead{TOI-6158 b}&
}
\startdata
\multicolumn{4}{l}{\hspace{-0.2cm} Orbital Parameters:}  \\
~~~$P$ & Orbital Period (Days) & 2.36712341$^{+0.0000033}_{-0.0000032}$ & 3.04468990$^{+0.00000551}_{-0.00000534}$ \\
~~~$e$ & Eccentricity & 0.045$^{+0.050}_{-0.032}$ & 0.059$^{+0.066}_{-0.042}$ \\
~~~$\omega$ & Argument of Periastron (Degrees) & -81$^{+190}_{-68}$ & 88$^{+57}_{-148}$ \\
~~~$K$ & Semi-amplitude Velocity (m/s) & 164 $\pm$ 20  & 95 $\pm$ 13 \\ 
~~~$\sigma_{\text{\tiny HPF}}$ & RV Jitter (m/s) & 19$^{+20}_{-13}$ & 32 $\pm$ 12 \\
\multicolumn{4}{l}{\hspace{-0.2cm} Transit Parameters:} \\
~~~$T_C$ & Transit Midpoint (BJD$_{\text{TDB}}$) & 2459817.6882 $\pm$ 0.0008 & 2459826.8508 $\pm$ 0.0008 \\
~~~$R_p / R_\star$ & Scaled Radius & 0.2213$^{+0.0080}_{-0.0080}$ & 0.2017$^{+0.0510}_{-0.0210}$ \\
~~~$a / R_\star$ & Scaled Semi-major Axis & 12.23$^{+0.35}_{-0.32}$ & 14.84$^{+0.35}_{-0.36}$ \\
~~~$i$ & Orbital Inclination (Degrees) & 89.13$^{+0.60}_{-0.69}$ & 86.51$^{+0.25}_{-0.32}$ \\
~~~$b$ & Impact Parameter & 0.19$^{+0.15}_{-0.13}$ & 0.86$^{+0.09}_{-0.06}$ \\
~~~$T_{14}$ & Transit Duration (Days) & 0.0740$^{+0.0020}_{-0.0024}$ & 0.0551$^{+0.0030}_{-0.0022}$ \\
~~~$\sigma_{\text{\tiny TESS Sector 55}}$ & Photometric Jitter (ppm) & 25$^{+193}_{-22}$ & -- \\
~~~$\sigma_{\text{\tiny TESS Sector 56}}$ & Photometric Jitter (ppm) & -- & 26$^{+227}_{-23}$ \\
~~~$\sigma_{\text{\tiny TESS Sector 82}}$ & Photometric Jitter (ppm) & 25.47$^{+223}_{-23}$ & 26$^{+236}_{-24}$ \\
~~~$\sigma_{\text{\tiny TESS Sector 83}}$ & Photometric Jitter (ppm) & -- & 31$^{+331}_{-28}$ \\
~~~$\sigma_{\text{\tiny RBO 20230903}}$ & Photometric Jitter (ppm) & 43$^{+512}_{-40}$ & -- \\
~~~$\sigma_{\text{\tiny Swope 20230730}}$ & Photometric Jitter (ppm) & -- & 41$^{+457}_{-38}$ \\
~~~$\sigma_{\text{\tiny Swope 20230811}}$ & Photometric Jitter (ppm) & -- & 48$^{+768}_{-45}$ \\
~~~$D_{\text{\tiny TESS 55}}$ & Dilution & 1.045$^{+0.083}_{-0.074}$ & -- \\
~~~$D_{\text{\tiny TESS 56}}$ & Dilution & -- & 1.475$^{+0.242}_{-0.213}$ \\
~~~$D_{\text{\tiny TESS 82}}$ & Dilution & 1.033$^{+0.074}_{-0.066}$ & 1.311$^{+0.220}_{-0.199}$ \\
~~~$D_{\text{\tiny TESS 83}}$ & Dilution & -- & 1.270$^{+0.214}_{-0.181}$ \\
\multicolumn{4}{l}{\hspace{-0.2cm} Planetary Parameters:}\\
~~~$M_p$ &  Mass ($M_{\oplus}$) & 219 $\pm$ 28 & 135$^{+19}_{-18}$ \\
~~~ & Mass ($M_{\text{J}}$) & 0.688$^{+0.087}_{-0.088}$ & 0.425$^{+0.060}_{-0.058}$ \\
~~~$R_p$ & Radius ($R_{\oplus}$) & 11.8$^{+0.52}_{-0.51}$ & 10.4$^{+2.70}_{-1.11}$ \\
~~~ & Radius ($R_{\text{J}}$) & 1.05 $\pm$ 0.05 & 0.93$^{+0.24}_{-0.10}$ \\
~~~$\rho_p$ & Density (g/cm$^3$) & 0.73$^{+0.14}_{-0.13}$ & 0.66$^{+0.41}_{-0.23}$ \\
~~~$a$ & Semi-major Axis (AU) & 0.02789 $\pm$ 0.00040 & 0.03275 $\pm$ 0.00040 \\
~~~$S$ & Planetary Insolation ($S_{\oplus}$) & 43.8 $\pm$ 3.8 & 27.3 $\pm$ 2.2 \\
~~~$T_{\text{\tiny eq}}$ & Equilibrium Temperature (K) & 716 $\pm$ 15 & 636 $\pm$ 13 \\
~~~$\langle F \rangle$ & Average Incident Flux ($10^4$ W/m$^2$) & 5.10 $\pm$ 0.58 & 3.70 $\pm$ 0.41 \\
\enddata
\tablecomments{An albedo of 0 is assumed}
\end{deluxetable*}

\subsection{Parameters from the spectral energy distribution}
We followed the procedures in \cite{canas_toi-3984_2023} to derive model-dependent stellar parameters with the \href{https://github.com/jdeast/EXOFASTv2}{\texttt{EXOFASTv2}} package \citep{Eastman2019}. \texttt{EXOFASTv2} models the observed spectral energy distribution (SED) using MIST models \citep{Dotter2016,Choi2016} and applies the $R_{v}=3.1$ reddening law from \cite{Fitzpatrick1999} to calculate a visual magnitude extinction. We placed Gaussian priors on the (i) broadband photometry listed in \autoref{tab:stellarparam}, (ii) spectroscopic parameters from \texttt{HPF-SpecMatch} (see Section \ref{sec:hpfspecmatch}), and (iii) parallax from Gaia DR3 \citep{GaiaCollaboration2022}. We placed a uniform prior on the visual extinction ($A_V$) with an upper limit derived from \cite{Green2019}. The stellar parameters are presented in  \autoref{tab:stellarparam}. TOI-5916 has a mass and radius of $M_\star=0.518\pm0.023~\mathrm{M_\odot}$ and $R_\star=0.487^{+0.013}_{-0.012}~\mathrm{R_\odot}$. TOI-6158 has a mass and radius of $M_\star=0.503^{+0.022}_{-0.023}~\mathrm{M_\odot}$ and $R_\star=0.476\pm0.013~\mathrm{R_\odot}$.

\section{Joint Fitting of Photometry and RVs} 
\label{sec: joint fit}

We use a Bayesian statistical modeling method implemented through the Python library \texttt{exoplanet} \citep{FM17} to estimate the orbital parameters of each planet. This library leverages PyMC3’s framework \citep{pymc3_Salvatier}, which uses Hamilton Monte Carlo (HMC) sampling. Using these tools, we jointly modeled the transit light curves and the RVs from various telescopes for both TOI-5916 and TOI-6158.
For both systems, we adopted a quadratic limb-darkening law, modeling independent limb-darkening coefficients for each instrument’s transit and sampling them following  \citet{Kipping2013}. For the TESS photometry, we included a dilution term to account for any unresolved background stars. 



For both TOI-5916 and TOI-6158, we did not use a Gaussian process (GP) to detrend the photometry because we did not see any evidence of correlated noise. When we included a simple harmonic oscillator GP, we found consistent results for the planetary radii regardless of the use of a GP.

Joint fit models were computed for each system twice, with and without a RV trend. For TOI-5916, we found a RV trend slope posterior value of $-2.06^{+93.81}_{-93.48}$ m/s/yr; for TOI-6158, we found a RV trend slope posterior value of $0.13^{+22.68}_{-23.50}$ m/s/yr. Since these slope values are centered around zero in both cases and their uncertainties are much larger than the slope values themselves, we found no statistically significant evidence for a non-zero RV trend posterior value. Additionally, in both cases, the values of $M_p$ and $K$ did not differ by any significant factor. Therefore, we report posterior results from the models that excluded RV trend terms to avoid over-complicating the models.

For TOI-5916, we found the dilution to be consistent with $1$. However, for TOI-6158, the TGLC pipeline may be slightly over-correcting, though the grazing geometry precludes us from giving a precise estimate on the dilution. Additionally, we incorporated a jitter term for HPF to account for any additional white noise not described by the formal measurement errors.



We show the transit model and phase-folded light curves in Figures \ref{fig:5916_PFLC} and \ref{fig:6158_PFLC}, and we show the best-fit RV models in Figures \ref{fig:5916_Combined_RV} and \ref{fig:6158_Combined_RV} for TOI-5916 b and TOI-6158 b, respectively. Table \ref{tab:PlanetParams} contains a summary of the derived planetary parameters. 


The transits of TOI-6158 b exhibit a distinctly ``V-shaped’’ morphology, indicative of a grazing geometry. Consistent with this, we measured a high impact parameter of 0.86$^{+0.12}_{-0.08}$.

\section{Discussion} \label{sec: discussion}
\subsection{TOI-5916 b and TOI-6158 b in the GEMS Parameter Space}

TOI-5916 b and TOI-6158 b join the growing list of about 33 other GEMS systems, contributing to the larger effort to characterize the overall population and constrain GEMS formation mechanisms, as outlined in the \textit{Searching for GEMS Survey} motivation \citep{motive}. The small sample size of GEMS systems allows us to begin to explore the processes responsible for the formation of these objects. Several trends have already begun to emerge from this sample of systems.

\begin{figure*}
    \centering
    \includegraphics[width=\textwidth]{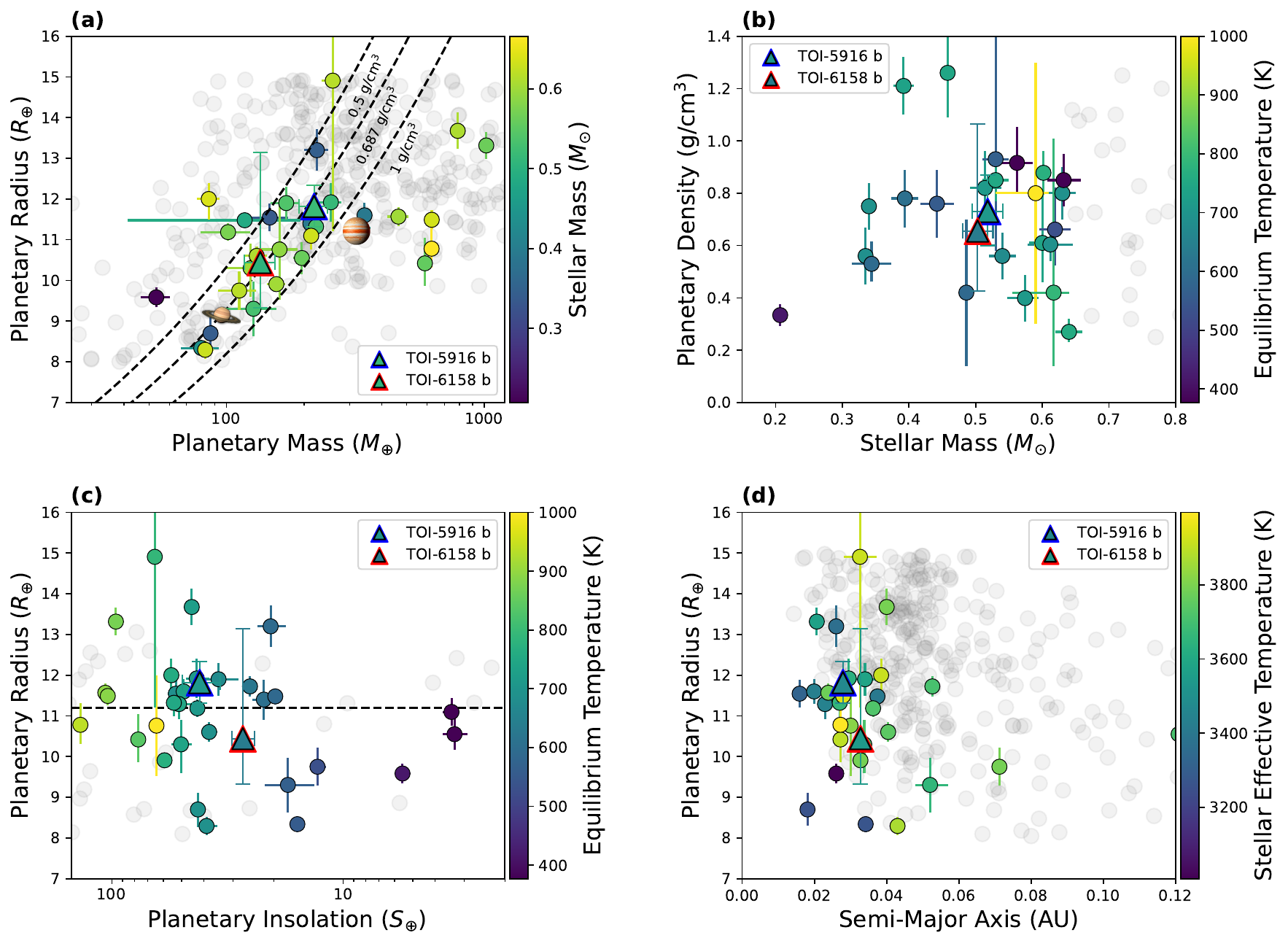}
    \caption{TOI-5916 and TOI-6158 are represented by the triangles highlighted in blue and red, respectively. The plots show the two systems, other GEMS, and FGK systems in various parameter spaces. Confirmed transiting GEMS systems are represented by the color-coded points. The gray points in the background are transiting giants around FGK dwarf stars. (a) The two target GEMS planets in the mass-radius parameter space. We include Saturn and Jupiter, as well as three lines of constant density, one being $0.5 \text{ g/cm}^3$, one being Saturn-density ($0.6873\text{ g/cm}^3$) and the other being water ($1 \text{ g/cm}^3$). Both TOI-5916 and TOI-6158 are close to the density of Saturn, as has been observed with prior GEMS systems. (b) The two target systems in the stellar mass-planetary density space. We see that, as expected, the planetary mass appears to decrease with the stellar mass. (c) The targets in the insolation-planetary radius parameter space. We do not see an upward trend in planetary radius with insolation, suggesting that these planets are not anomalously inflated. For reference, the radius of Jupiter is also indicated by the black dashed line. (d) The targets in the semi-major axis-planetary radius space. We see that there is also no upward trend with planetary radius of semi-major axis.}
    \label{fig:Parameter_Plots}
\end{figure*}

To better observe these trends, in Figure \ref{fig:Parameter_Plots}, we plot TOI-5916 b and TOI-6158 b against the other transiting GEMS systems in various parameter spaces.  We compare the systems with other GEMS and transiting giant exoplanets orbiting FGK dwarfs. We obtained the planet parameters from the NASA Exoplanet Archive \citep{NASAArchive_Paper, NASA_EXOARCH} where the following filters were used: $ 8\earthradius \lesssim R_p \lesssim 15\earthradius $, $ 2600K <  T_{\text{eff}} < 4000K $, and $0.08\solmass < M_* < 0.6\solmass$. This query results in a final sample size of 34 transiting GEMS systems that were used in our plots. The FGK dwarf systems were selected using the following criteria: $ 8\earthradius \lesssim R_p \lesssim 15\earthradius $, $ 0.7\solradius < R* < 1.6\solradius $, $ 3700K <  T_{\text{eff}} < 7500K $, and $ 0.6\solmass < M_* < 1.4\solmass $. All data for this comparison were obtained on 2025 September 3. In addition to these GEMS systems listed on the NASA Exoplanet Archive, we added two recently published GEMS systems that are not yet available on the archive. These systems are TOI-6303 b \citep{Hotnisky_2025} and TOI-6330 b \citep{Hotnisky_2025}, making the total count of GEMS systems now 35.

For 8 of the GEMS systems, a reliable insolation value was not available from the NASA Exoplanet Archive. In order to remedy this issue, the insolation value (and its associated errors) was calculated independently using the parameters available from the archive. This was done by checking for \texttt{NaN} insolation values and applying the following formula, where $S$ is the insolation:

\begin{equation}
    S = S_{\oplus} \left(\frac{R_{\star}}{R_{\odot}}\right)^2 \left(\frac{T_{\text{eff}}}{T_{\odot}}\right)^4 \left(\frac{a_\oplus}{a}\right)^2
\end{equation}

Note that $R_{\star}$ is in units of $R_{\odot}$, $T_{\text{eff}}$ is in units of K, and $a$ is in units of AU. Standard error propagation was also done to account for the errors of those insolation values. This calculation allowed us to obtain insolation values for all GEMS systems, which are shown in Figure \ref{fig:Parameter_Plots}c. Note that the plotted values are expressed in units of Earth's insolation, $S_{\oplus}$.


In the cases of TOI-5916 b ($\rho_\text{5916} = 0.73^{+0.14}_{-0.13} \text{ g/cm}^3$) and TOI-6158 b ($\rho_\text{6158} =0.66^{+0.41}_{-0.23} \text{ g/cm}^3$), these systems add to the growing trend that GEMS are Saturn-density. This is best illustrated in Figure \ref{fig:Parameter_Plots}a, where we can see that most transiting GEMS (20/35) fall within the $0.5$ g cm$^{-3}$ to $1$ g cm$^{-3}$ density envelope. However, as stated in \cite{motive}, about 5 more GEMS systems will need to be characterized to ensure that this is not just the result of the relatively small sample size. 

We find a trend in the orbital semi-major axes of the GEMS systems (Figure \ref{fig:Parameter_Plots}d), which tend to lie at closer separations than those of giant planets orbiting FGK dwarfs. Part of this effect could arise from observational bias, since the smaller sizes of M dwarfs reduce the transit probability of a planet at the same orbital distance compared to FGK dwarfs. In Section \ref{sec: semi-major}, we test whether this difference persists after correcting for the transit probability.

In Figure \ref{fig:Parameter_Plots}c, there is not a significant upward trend in the radius with insolation, suggesting that GEMS are generally not being inflated by their host stars' energy. This is expected, as GEMS' cool-M-dwarf hosts are not hot enough to inflate their planetary companions, thus GEMS do not exhibit the high equilibrium temperatures and inflated radii of hot Jupiters despite their being in typical ``hot-Jupiter" period space.


\subsection{Two new Saturn-density GEMS}
The maximum masses of giant planets have been observed to decrease with stellar mass \citep[][]{Kanodia2025}, corresponding to the unpopulated region in the upper left of Figure \ref{fig:Parameter_Plots}d. TOI-5916 b and TOI-6158 b follow this trend, exhibiting typical masses relative to the masses of their host stars. More interestingly, they both join the majority of GEMS in density space, having a density similar to Saturn. The grazing transit of TOI-6158 b has made it difficult to measure its precise planet radius, but it does have an $\sim84\%$ probability to be less dense than water (Figure \ref{fig:Parameter_Plots}a). 

The scarcity of super-Jupiters and the abundance in Saturn-density planets around M-dwarf stars are among the most striking characteristics of GEMS. The former may be explained by the lower average dust masses in the M-dwarf protoplanetary disks and slower accretion timescales under the core accretion paradigm \citep{Laughlin}. Although the lack of super-giants around M-dwarfs may be expected, the formation mechanism of these Saturn-density GEMS remains an open question. Interestingly, these Saturn-density giant planets may not be only orbiting M dwarfs: excluding planets heavier than 2 $M_J$, the remaining warm Jupiters around FGK dwarfs also seem to populate near Saturn density \citep{Kanodia2025b}. In other words, these Saturn-density warm giant planets around FGK+M dwarfs may share a similar formation pathway independent of host star mass. The addition of these two GEMS strengthens this finding, and may help us to understand their formation as part of a broader mechanism shaping transiting warm, Saturn-density gas giants across a wide range of stellar masses.

\subsection{Semi-major Axis Distribution of Giant Planets} \label{sec: semi-major}
\begin{figure*}
    \centering
    \includegraphics[width=\textwidth]{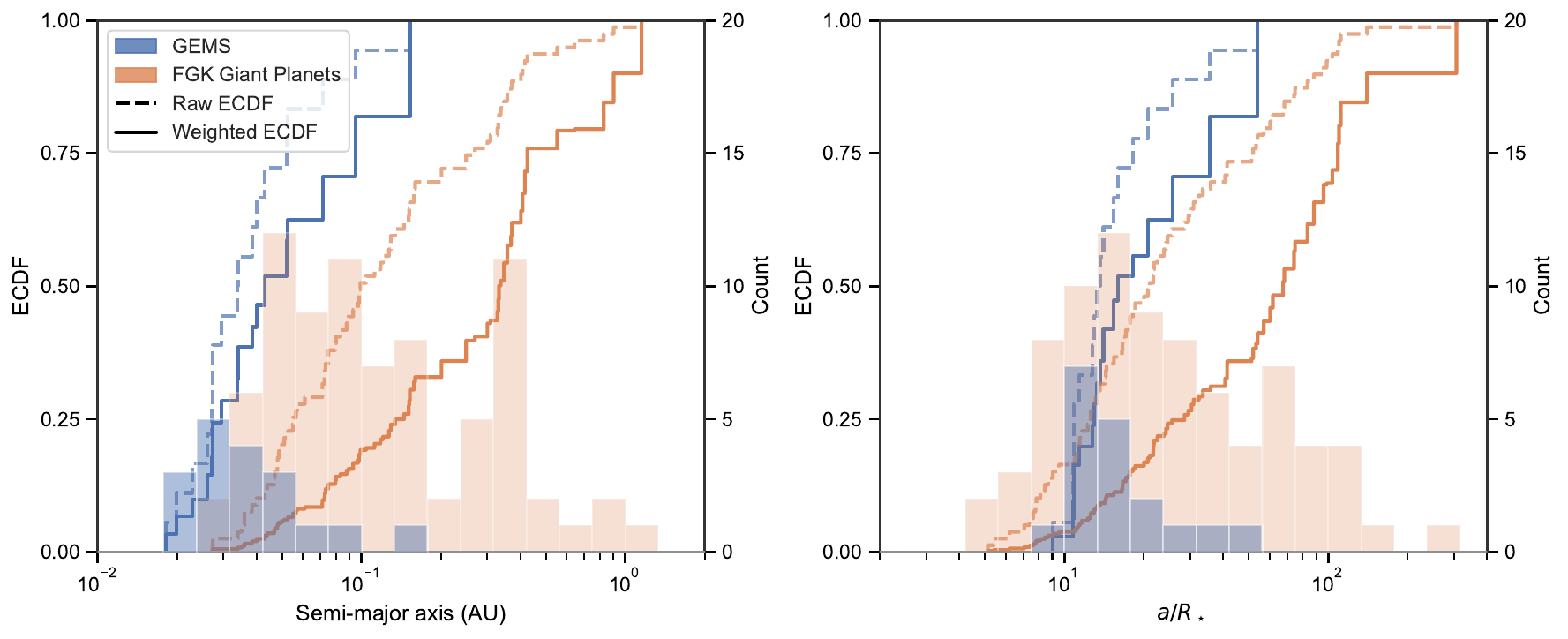}
    \caption{Distributions of semi-major axis (left) and scaled semi-major axis, $a/R_\star$ (right), for giant planets around M dwarfs (blue) and FGK dwarfs (orange) discovered by \textit{TESS}. The histograms show the raw distribution of semi-major axes. Dotted curves show the observed empirical cumulative distribution functions (ECDFs) of these raw distribution, while solid curves show the bias-corrected ECDFs obtained by weighting each system by the inverse of its geometric transit probability. Although the GEMS sample is small, the bias-corrected ECDFs still suggest that giant planets around M dwarfs preferentially occupy closer orbits than their counterparts around FGK dwarfs. }
    \label{fig:ecdf}
\end{figure*}
We further explore whether there exists a trend for GEMS to have closer orbits than FGK giants. Most transiting GEMS have been discovered by TESS, naturally biasing towards planets with orbital periods under $\sim 14$ days. Therefore, we filtered both populations to contain only planets discovered by TESS. 

To account for the geometric selection effects that bias transit detections toward close-in planets, we applied a transit probability correction to the GEMS and FGK hot Jupiter samples. The probability that a planet with semi-major axis $a$ and eccentricity $e$ transits a star of radius $R_\star$ is given by
\begin{equation}
    P_{\rm t} = \frac{R_\star + R_p}{a} \, \frac{1 + e \sin \omega}{1 - e^2},
\end{equation}
where $R_p$ is the planetary radius and $\omega$ is the argument of periastron. For circular orbits, this expression reduces to $(R_\star + R_p)/a$. We calculated $P_{\rm t}$ for each planet using the reported stellar and planetary parameters. Each system was then assigned a weight $w \propto 1/P_{\rm t}$, following the Horvitz--Thompson prescription \citep{Horvitz_1952}, such that intrinsically less probable transits contribute more heavily to the inferred distribution. Within each stellar population, the weights were normalized so that the cumulative distributions integrate to unity. We then constructed weighted empirical cumulative distribution functions (ECDFs) to compare the intrinsic orbital architectures before and after applying the correction (Figure \ref{fig:ecdf}).

The ECDF of GEMS remains shifted toward smaller semi-major axes compared to giant planets around FGK dwarfs, even after correcting for transit probability. The bias-corrected weighted median semi-major axis for GEMS is $a_{\rm GEMS} = 0.042^{+0.060}_{-0.015}$ AU, while for FGK dwarfs it is $a_{\rm FGK} = 0.33^{+0.47}_{-0.31}$ AU. A similar offset is evident in the scaled semi-major axis, reinforcing that the difference is unlikely to arise from observational bias against smaller host stars (Figure \ref{fig:ecdf}). Although we need a larger GEMS sample to statistically confirm this trend, the tendency toward closer orbits may arise from the lower masses of M dwarfs, which require closer orbits to initiate tidal circularization following high-eccentricity migration. 

\section{Conclusion} \label{sec: conclusion}
We report the discovery and confirmation of TOI-5916~b and TOI-6158~b as part of the \textit{Searching for GEMS} Survey. Both giant planets are Saturn-density and have radii comparable to that of Jupiter. Both systems also orbit M2 dwarfs with very similar spectroscopic properties. 

TOI-5916 b and TOI-6158 b were both discovered with TESS and then confirmed with RV follow-up from HPF and photometric follow-up from other ground-based sources. The photometric ground-based data used to confirm TOI-5916 b were from RBO, while the photometric ground-based data used to confirm TOI-6158 b were from Swope at Las Campanas.

These two systems join the ever-growing group of 32 other GEMS systems: TOI-5916 b with a mass of $219\pm 28~M_{\oplus}$, radius of $11.8^{+0.52}_{-0.51}~R_{\oplus}$ and density of $0.73^{+0.14}_{-0.13}$ g/cm$^3$, and TOI-6158 b with a mass of $135^{+19}_{-18}~M_{\oplus}$, radius of $10.4^{+2.70}_{-1.11}~R_{\oplus}$, and density of 0.66$^{+0.41}_{-0.23}$ g/cm$^3$.

By comparing TOI-5916 b and TOI-6158 b to other confirmed GEMS systems and FGK giants in various different parameter spaces, we were able to show how these two systems help to further our understanding of GEMS. Through this work, we were able to further show that GEMS' systems tend to follow these trends: i) transiting GEMS appear to tend towards a Saturn-like density. ii.) GEMS appear to tend towards shorter, sub-four day periods, though more data are needed to confirm this due to bias of transiting systems. iii.) There does not appear to be a significant upward trend in radius with insolation, suggesting that GEMS are not inflated by host star energy. iv). The maximum masses of transiting giant planets tend to decrease with stellar mass.

Finally, we note that the addition of these two GEMS system to the \textit{Searching for GEMS Survey} helps to strengthen this notion that Saturn-density warm Jupiters around FGKM dwarfs may share a similar formation pathway independent of stellar mass.



\section{Acknowledgments}
These results are based on observations obtained with the Habitable-zone Planet Finder Spectrograph on the HET. The HPF team acknowledges support from NSF grants AST-1006676, AST-1126413, AST-1310885, AST-1517592, AST-1310875, ATI 2009889, ATI-2009982, AST-2108512, ATI 2009554, and the NASA Astrobiology Institute (NNA09DA76A) in the pursuit of precision radial velocities in the NIR. The HPF team also acknowledges support from the Heising-Simons Foundation via grant 2017-0494.  

This study was supported by NASA grant 80NSSC25K0184 as part of the U.S. Contributions to Ariel Preparatory Science (USCAPS) Program.

Based on observations obtained with the Hobby-Eberly Telescope (HET), which is a joint project of the University of Texas at Austin, the Pennsylvania State University, Ludwig-Maximillians-Universitaet Muenchen, and Georg-August Universitaet Goettingen. The HET is named in honor of its principal benefactors, William P. Hobby and Robert E. Eberly.  We wish to thank the HET night operations staff for their assistance in the collection of data used herein.  We acknowledge the Texas Advanced Computing Center (TACC) at The University of Texas at Austin for providing high performance computing, visualization, and storage resources that have contributed to the results reported within this paper.

We would like to acknowledge that the HET is built on Indigenous land. Moreover, we would like to acknowledge and pay our respects to the Carrizo \& Comecrudo, Coahuiltecan, Caddo, Tonkawa, Comanche, Lipan Apache, Alabama-Coushatta, Kickapoo, Tigua Pueblo, and all the American Indian and Indigenous Peoples and communities who have been or have become a part of these lands and territories in Texas, on Turtle Island.

Some of the observations in this paper made use of the NN-EXPLORE Exoplanet and Stellar Speckle Imager (NESSI). NESSI was funded by the NASA Exoplanet Exploration Program and the NASA Ames Research Center. NESSI was built at the Ames Research Center by Steve B. Howell, Nic Scott, Elliott P. Horch, and Emmett Quigley (NN-EXPLORE Program ID: 2023B-438370).

Some of the data presented in this paper were obtained from
MAST at STScI. Support for MAST for non-HST data is
provided by the NASA Office of Space Science via Grant
NNX09AF08G and by other grants and contracts. This work
includes data collected by the TESS mission, which are
publicly available from MAST. Funding for the TESS mission
is provided by the NASA Science Mission Directorate.

CIC acknowledges support by NASA Headquarters through an appointment to the NASA Postdoctoral Program at the Goddard Space Flight Center, administered by ORAU through a contract with NASA. 

We thank Dr. Margaret Wright for her assistance with proofreading this manuscript.

\clearpage
\bibliography{references}{}

\begin{thebibliography}{}
\expandafter\ifx\csname natexlab\endcsname\relax\def\natexlab#1{#1}\fi
\providecommand{\url}[1]{\href{#1}{#1}}
\providecommand{\dodoi}[1]{doi:~\href{http://doi.org/#1}{\nolinkurl{#1}}}
\providecommand{\doeprint}[1]{\href{http://ascl.net/#1}{\nolinkurl{http://ascl.net/#1}}}
\providecommand{\doarXiv}[1]{\href{https://arxiv.org/abs/#1}{\nolinkurl{https://arxiv.org/abs/#1}}}

\bibitem[{{Akeson} {et~al.}(2013){Akeson}, {Chen}, {Ciardi}, {Crane}, {Good}, {Harbut}, {Jackson}, {Kane}, {Laity}, {Leifer}, {Lynn}, {McElroy}, {Papin}, {Plavchan}, {Ram{\'\i}rez}, {Rey}, {von Braun}, {Wittman}, {Abajian}, {Ali}, {Beichman}, {Beekley}, {Berriman}, {Berukoff}, {Bryden}, {Chan}, {Groom}, {Lau}, {Payne}, {Regelson}, {Saucedo}, {Schmitz}, {Stauffer}, {Wyatt}, \& {Zhang}}]{NASAArchive_Paper}
{Akeson}, R.~L., {Chen}, X., {Ciardi}, D., {et~al.} 2013, \pasp, 125, 989, \dodoi{10.1086/672273}

\bibitem[{Boss(2006)}]{Boss06}
Boss, A.~P. 2006, The Astrophysical Journal, 641, 1148, \dodoi{10.1086/500530}

\bibitem[{{Boss} \& {Kanodia}(2023)}]{Boss2023}
{Boss}, A.~P., \& {Kanodia}, S. 2023, \apj, 956, 4, \dodoi{10.3847/1538-4357/acf373}

\bibitem[{{Bryant} {et~al.}(2023){Bryant}, {Bayliss}, \& {Van Eylen}}]{Bryant_2023}
{Bryant}, E.~M., {Bayliss}, D., \& {Van Eylen}, V. 2023, \mnras, 521, 3663, \dodoi{10.1093/mnras/stad626}

\bibitem[{{Bryant} {et~al.}(2024){Bryant}, {Bayliss}, {Hartman}, {Sedaghati}, {Hobson}, {Jord{\'a}n}, {Brahm}, {Bakos}, {Almenara}, {Barkaoui}, {Bonfils}, {Cointepas}, {Collins}, {Dransfield}, {Evans}, {Gillon}, {Jehin}, {Murgas}, {Pozuelos}, {Schwarz}, {Timmermans}, {Watkins}, {W{\"u}nsche}, {Butler}, {Crane}, {Shectman}, {Teske}, {Charbonneau}, {Essack}, {Jenkins}, {Lewis}, {Seager}, {Ting}, \& {Winn}}]{Bryant2024}
{Bryant}, E.~M., {Bayliss}, D., {Hartman}, J.~D., {et~al.} 2024, \mnras, 533, 3893, \dodoi{10.1093/mnras/stae2034}

\bibitem[{{Burn} {et~al.}(2021){Burn}, {Schlecker}, {Mordasini}, {Emsenhuber}, {Alibert}, {Henning}, {Klahr}, \& {Benz}}]{Burn_2021}
{Burn}, R., {Schlecker}, M., {Mordasini}, C., {et~al.} 2021, \aap, 656, A72, \dodoi{10.1051/0004-6361/202140390}

\bibitem[{{Ca{\~n}as} {et~al.}(2020){Ca{\~n}as}, {Stefansson}, {Kanodia}, {Mahadevan}, {Cochran}, {Endl}, {Robertson}, {Bender}, {Ninan}, {Beard}, {Lubin}, {Gupta}, {Everett}, {Monson}, {Wilson}, {Lewis}, {Brewer}, {Majewski}, {Hebb}, {Dawson}, {Diddams}, {Ford}, {Fredrick}, {Halverson}, {Hearty}, {Lin}, {Metcalf}, {Rajagopal}, {Ramsey}, {Roy}, {Schwab}, {Terrien}, \& {Wright}}]{Canas2020}
{Ca{\~n}as}, C.~I., {Stefansson}, G., {Kanodia}, S., {et~al.} 2020, \aj, 160, 147, \dodoi{10.3847/1538-3881/abac67}

\bibitem[{{Ca{\~n}as} {et~al.}(2023){Ca{\~n}as}, {Kanodia}, {Libby-Roberts}, {Lin}, {Schutte}, {Powers}, {Jones}, {Monson}, {Wang}, {Stef{\'a}nsson}, {Cochran}, {Robertson}, {Mahadevan}, {Kowalski}, {Wisniewski}, {Parker}, {Larsen}, {Chapman}, {Kobulnicky}, {Gupta}, {Everett}, {Penprase}, {Zeimann}, {Beard}, {Bender}, {Col{\'o}n}, {Diddams}, {Fredrick}, {Halverson}, {Ninan}, {Ramsey}, {Roy}, \& {Schwab}}]{canas_toi-3984_2023}
{Ca{\~n}as}, C.~I., {Kanodia}, S., {Libby-Roberts}, J., {et~al.} 2023, \aj, 166, 30, \dodoi{10.3847/1538-3881/acdac7}

\bibitem[{{Chambers} {et~al.}(2016){Chambers}, {Magnier}, {Metcalfe}, {Flewelling}, {Huber}, {Waters}, {Denneau}, {Draper}, {Farrow}, {Finkbeiner}, {Holmberg}, {Koppenhoefer}, {Price}, {Rest}, {Saglia}, {Schlafly}, {Smartt}, {Sweeney}, {Wainscoat}, {Burgett}, {Chastel}, {Grav}, {Heasley}, {Hodapp}, {Jedicke}, {Kaiser}, {Kudritzki}, {Luppino}, {Lupton}, {Monet}, {Morgan}, {Onaka}, {Shiao}, {Stubbs}, {Tonry}, {White}, {Ba{\~n}ados}, {Bell}, {Bender}, {Bernard}, {Boegner}, {Boffi}, {Botticella}, {Calamida}, {Casertano}, {Chen}, {Chen}, {Cole}, {Deacon}, {Frenk}, {Fitzsimmons}, {Gezari}, {Gibbs}, {Goessl}, {Goggia}, {Gourgue}, {Goldman}, {Grant}, {Grebel}, {Hambly}, {Hasinger}, {Heavens}, {Heckman}, {Henderson}, {Henning}, {Holman}, {Hopp}, {Ip}, {Isani}, {Jackson}, {Keyes}, {Koekemoer}, {Kotak}, {Le}, {Liska}, {Long}, {Lucey}, {Liu}, {Martin}, {Masci}, {McLean}, {Mindel}, {Misra}, {Morganson}, {Murphy}, {Obaika}, {Narayan}, {Nieto-Santisteban}, {Norberg}, {Peacock}, {Pier}, {Postman}, {Primak}, {Rae}, {Rai},
  {Riess}, {Riffeser}, {Rix}, {R{\"o}ser}, {Russel}, {Rutz}, {Schilbach}, {Schultz}, {Scolnic}, {Strolger}, {Szalay}, {Seitz}, {Small}, {Smith}, {Soderblom}, {Taylor}, {Thomson}, {Taylor}, {Thakar}, {Thiel}, {Thilker}, {Unger}, {Urata}, {Valenti}, {Wagner}, {Walder}, {Walter}, {Watters}, {Werner}, {Wood-Vasey}, \& {Wyse}}]{PS1_DR2}
{Chambers}, K.~C., {Magnier}, E.~A., {Metcalfe}, N., {et~al.} 2016, arXiv e-prints, arXiv:1612.05560, \dodoi{10.48550/arXiv.1612.05560}

\bibitem[{{Choi} {et~al.}(2016){Choi}, {Dotter}, {Conroy}, {Cantiello}, {Paxton}, \& {Johnson}}]{Choi2016}
{Choi}, J., {Dotter}, A., {Conroy}, C., {et~al.} 2016, \apj, 823, 102, \dodoi{10.3847/0004-637X/823/2/102}

\bibitem[{{Collins} {et~al.}(2017){Collins}, {Kielkopf}, {Stassun}, \& {Hessman}}]{KACollins}
{Collins}, K.~A., {Kielkopf}, J.~F., {Stassun}, K.~G., \& {Hessman}, F.~V. 2017, \aj, 153, 77, \dodoi{10.3847/1538-3881/153/2/77}

\bibitem[{{Cutri} {et~al.}(2003){Cutri}, {Skrutskie}, {van Dyk}, {Beichman}, {Carpenter}, {Chester}, {Cambresy}, {Evans}, {Fowler}, {Gizis}, {Howard}, {Huchra}, {Jarrett}, {Kopan}, {Kirkpatrick}, {Light}, {Marsh}, {McCallon}, {Schneider}, {Stiening}, {Sykes}, {Weinberg}, {Wheaton}, {Wheelock}, \& {Zacarias}}]{2MASS}
{Cutri}, R.~M., {Skrutskie}, M.~F., {van Dyk}, S., {et~al.} 2003, {VizieR Online Data Catalog: 2MASS All-Sky Catalog of Point Sources (Cutri+ 2003)}, VizieR On-line Data Catalog: II/246. Originally published in: University of Massachusetts and Infrared Processing and Analysis Center, (IPAC/California Institute of Technology) (2003)

\bibitem[{{Cutri} {et~al.}(2021){Cutri}, {Wright}, {Conrow}, {Fowler}, {Eisenhardt}, {Grillmair}, {Kirkpatrick}, {Masci}, {McCallon}, {Wheelock}, {Fajardo-Acosta}, {Yan}, {Benford}, {Harbut}, {Jarrett}, {Lake}, {Leisawitz}, {Ressler}, {Stanford}, {Tsai}, {Liu}, {Helou}, {Mainzer}, {Gettngs}, {Gonzalez}, {Hoffman}, {Marsh}, {Padgett}, {Skrutskie}, {Beck}, {Papin}, \& {Wittman}}]{WISE}
{Cutri}, R.~M., {Wright}, E.~L., {Conrow}, T., {et~al.} 2021, {VizieR Online Data Catalog: AllWISE Data Release (Cutri+ 2013)}, VizieR On-line Data Catalog: II/328. Originally published in: IPAC/Caltech (2013)

\bibitem[{{Dong} {et~al.}(2025){Dong}, {Chontos}, {Zhou}, {Stefansson}, {Wang}, {Huang}, {Gupta}, {Halverson}, {Kanodia}, {Luhn}, {Mahadevan}, {Monson}, {Alvarado-Montes}, {Ninan}, {Robertson}, {Roy}, {Schwab}, \& {Wright}}]{Dong2025}
{Dong}, J., {Chontos}, A., {Zhou}, G., {et~al.} 2025, \aj, 169, 4, \dodoi{10.3847/1538-3881/ad8ec1}

\bibitem[{{Dotter}(2016)}]{Dotter2016}
{Dotter}, A. 2016, \apjs, 222, 8, \dodoi{10.3847/0067-0049/222/1/8}

\bibitem[{{Eastman} {et~al.}(2019){Eastman}, {Rodriguez}, {Agol}, {Stassun}, {Beatty}, {Vanderburg}, {Gaudi}, {Collins}, \& {Luger}}]{Eastman2019}
{Eastman}, J.~D., {Rodriguez}, J.~E., {Agol}, E., {et~al.} 2019, arXiv e-prints, arXiv:1907.09480.
\newblock \doarXiv{1907.09480}

\bibitem[{{Fitzpatrick}(1999)}]{Fitzpatrick1999}
{Fitzpatrick}, E.~L. 1999, \pasp, 111, 63, \dodoi{10.1086/316293}

\bibitem[{Foreman-Mackey {et~al.}(2017)Foreman-Mackey, Agol, Ambikasaran, \& Angus}]{FM17}
Foreman-Mackey, D., Agol, E., Ambikasaran, S., \& Angus, R. 2017, The Astronomical Journal, 154, 220, \dodoi{10.3847/1538-3881/aa9332}

\bibitem[{{Gaia Collaboration} {et~al.}(2022){Gaia Collaboration}, {Vallenari, A.}, {Brown, A.G.A.}, {Prusti, T.}, \& {et al.}}]{GaiaCollaboration2022}
{Gaia Collaboration}, {Vallenari, A.}, {Brown, A.G.A.}, {Prusti, T.}, \& {et al.} 2022, A\&A, \dodoi{10.1051/0004-6361/202243940}

\bibitem[{{Gaia Collaboration} {et~al.}(2023){Gaia Collaboration}, {Vallenari}, {Brown}, {Prusti}, {de Bruijne}, {Arenou}, {Babusiaux}, {Biermann}, {Creevey}, {Ducourant}, {Evans}, {Eyer}, {Guerra}, {Hutton}, {Jordi}, {Klioner}, {Lammers}, {Lindegren}, {Luri}, {Mignard}, {Panem}, {Pourbaix}, {Randich}, {Sartoretti}, {Soubiran}, {Tanga}, {Walton}, {Bailer-Jones}, {Bastian}, {Drimmel}, {Jansen}, {Katz}, {Lattanzi}, {van Leeuwen}, {Bakker}, {Cacciari}, {Casta{\~n}eda}, {De Angeli}, {Fabricius}, {Fouesneau}, {Fr{\'e}mat}, {Galluccio}, {Guerrier}, {Heiter}, {Masana}, {Messineo}, {Mowlavi}, {Nicolas}, {Nienartowicz}, {Pailler}, {Panuzzo}, {Riclet}, {Roux}, {Seabroke}, {Sordo}, {Th{\'e}venin}, {Gracia-Abril}, {Portell}, {Teyssier}, {Altmann}, {Andrae}, {Audard}, {Bellas-Velidis}, {Benson}, {Berthier}, {Blomme}, {Burgess}, {Busonero}, {Busso}, {C{\'a}novas}, {Carry}, {Cellino}, {Cheek}, {Clementini}, {Damerdji}, {Davidson}, {de Teodoro}, {Nu{\~n}ez Campos}, {Delchambre}, {Dell'Oro}, {Esquej},
  {Fern{\'a}ndez-Hern{\'a}ndez}, {Fraile}, {Garabato}, {Garc{\'\i}a-Lario}, {Gosset}, {Haigron}, {Halbwachs}, {Hambly}, {Harrison}, {Hern{\'a}ndez}, {Hestroffer}, {Hodgkin}, {Holl}, {Jan{\ss}en}, {Jevardat de Fombelle}, {Jordan}, {Krone-Martins}, {Lanzafame}, {L{\"o}ffler}, {Marchal}, {Marrese}, {Moitinho}, {Muinonen}, {Osborne}, {Pancino}, {Pauwels}, {Recio-Blanco}, {Reyl{\'e}}, {Riello}, {Rimoldini}, {Roegiers}, {Rybizki}, {Sarro}, {Siopis}, {Smith}, {Sozzetti}, {Utrilla}, {van Leeuwen}, {Abbas}, {{\'A}brah{\'a}m}, {Abreu Aramburu}, {Aerts}, {Aguado}, {Ajaj}, {Aldea-Montero}, {Altavilla}, {{\'A}lvarez}, {Alves}, {Anders}, {Anderson}, {Anglada Varela}, {Antoja}, {Baines}, {Baker}, {Balaguer-N{\'u}{\~n}ez}, {Balbinot}, {Balog}, {Barache}, {Barbato}, {Barros}, {Barstow}, {Bartolom{\'e}}, {Bassilana}, {Bauchet}, {Becciani}, {Bellazzini}, {Berihuete}, {Bernet}, {Bertone}, {Bianchi}, {Binnenfeld}, {Blanco-Cuaresma}, {Blazere}, {Boch}, {Bombrun}, {Bossini}, {Bouquillon}, {Bragaglia}, {Bramante}, {Breedt},
  {Bressan}, {Brouillet}, {Brugaletta}, {Bucciarelli}, {Burlacu}, {Butkevich}, {Buzzi}, {Caffau}, {Cancelliere}, {Cantat-Gaudin}, {Carballo}, {Carlucci}, {Carnerero}, {Carrasco}, {Casamiquela}, {Castellani}, {Castro-Ginard}, {Chaoul}, {Charlot}, {Chemin}, {Chiaramida}, {Chiavassa}, {Chornay}, {Comoretto}, {Contursi}, {Cooper}, {Cornez}, {Cowell}, {Crifo}, {Cropper}, {Crosta}, {Crowley}, {Dafonte}, {Dapergolas}, {David}, {David}, {de Laverny}, {De Luise}, \& {De March}}]{Gaia_2023}
{Gaia Collaboration}, {Vallenari}, A., {Brown}, A.~G.~A., {et~al.} 2023, \aap, 674, A1, \dodoi{10.1051/0004-6361/202243940}

\bibitem[{{Gan} {et~al.}(2025){Gan}, {Theissen}, {Wang}, {Burgasser}, \& {Mao}}]{Gan2025}
{Gan}, T., {Theissen}, C.~A., {Wang}, S.~X., {Burgasser}, A.~J., \& {Mao}, S. 2025, \apjs, 276, 47, \dodoi{10.3847/1538-4365/ad9c65}

\bibitem[{Gan {et~al.}(2022)Gan, Wang, Wang, Mao, Huang, Collins, Stassun, Shporer, Zhu, Ricker, Vanderspek, Latham, Seager, Winn, Jenkins, Barkaoui, Belinski, Ciardi, Evans, Girardin, Maslennikova, Mazeh, Panahi, Pozuelos, Radford, Schwarz, Twicken, Wünsche, \& Zucker}]{Gan_23}
Gan, T., Wang, S.~X., Wang, S., {et~al.} 2022, The Astronomical Journal, 165, 17, \dodoi{10.3847/1538-3881/ac9b12}

\bibitem[{{Gardner} {et~al.}(2006){Gardner}, {Mather}, {Clampin}, {Doyon}, {Greenhouse}, {Hammel}, {Hutchings}, {Jakobsen}, {Lilly}, {Long}, {Lunine}, {McCaughrean}, {Mountain}, {Nella}, {Rieke}, {Rieke}, {Rix}, {Smith}, {Sonneborn}, {Stiavelli}, {Stockman}, {Windhorst}, \& {Wright}}]{JWST}
{Gardner}, J.~P., {Mather}, J.~C., {Clampin}, M., {et~al.} 2006, \ssr, 123, 485, \dodoi{10.1007/s11214-006-8315-7}

\bibitem[{{Green} {et~al.}(2019){Green}, {Schlafly}, {Zucker}, {Speagle}, \& {Finkbeiner}}]{Green2019}
{Green}, G.~M., {Schlafly}, E., {Zucker}, C., {Speagle}, J.~S., \& {Finkbeiner}, D. 2019, \apj, 887, 93, \dodoi{10.3847/1538-4357/ab5362}

\bibitem[{Han \& Brandt(2023)}]{Han_2023}
Han, T., \& Brandt, T.~D. 2023, The Astronomical Journal, 165, 71, \dodoi{10.3847/1538-3881/acaaa7}

\bibitem[{{Han} {et~al.}(2025){Han}, {Robertson}, {Brandt}, {Kanodia}, {Ca{\~n}as}, {Shporer}, {Ricker}, \& {Beard}}]{Han2025}
{Han}, T., {Robertson}, P., {Brandt}, T.~D., {et~al.} 2025, \apjl, 988, L4, \dodoi{10.3847/2041-8213/ade794}

\bibitem[{{Han} {et~al.}(2024){Han}, {Robertson}, {Kanodia}, {Ca{\~n}as}, {Lin}, {Stef{\'a}nsson}, {Libby-Roberts}, {Larsen}, {Kobulnicky}, {Mahadevan}, {Bender}, {Cochran}, {Endl}, {Everett}, {Gupta}, {Halverson}, {Hearty}, {Monson}, {Ninan}, {Roy}, {Schwab}, \& {Terrien}}]{Han2024}
{Han}, T., {Robertson}, P., {Kanodia}, S., {et~al.} 2024, \aj, 167, 4, \dodoi{10.3847/1538-3881/ad09c2}

\bibitem[{{Hill} {et~al.}(2021){Hill}, {Lee}, {MacQueen}, {Kelz}, {Drory}, {Vattiat}, {Good}, {Ramsey}, {Kriel}, {Peterson}, {DePoy}, {Gebhardt}, {Marshall}, {Tuttle}, {Bauer}, {Chonis}, {Fabricius}, {Froning}, {H{\"a}user}, {Indahl}, {Jahn}, {Landriau}, {Leck}, {Montesano}, {Prochaska}, {Snigula}, {Zeimann}, {Bryant}, {Damm}, {Fowler}, {Janowiecki}, {Martin}, {Mrozinski}, {Odewahn}, {Rostopchin}, {Shetrone}, {Spencer}, {Mentuch Cooper}, {Armandroff}, {Bender}, {Dalton}, {Hopp}, {Komatsu}, {Nicklas}, {Ramsey}, {Roth}, {Schneider}, {Sneden}, \& {Steinmetz}}]{Hill2021}
{Hill}, G.~J., {Lee}, H., {MacQueen}, P.~J., {et~al.} 2021, \aj, 162, 298, \dodoi{10.3847/1538-3881/ac2c02}

\bibitem[{Horvitz \& Thompson(1952)}]{Horvitz_1952}
Horvitz, D.~G., \& Thompson, D.~J. 1952, Journal of the American Statistical Association, 47, 663, \dodoi{10.1080/01621459.1952.10483446}

\bibitem[{{Hotnisky} {et~al.}(2025){Hotnisky}, {Kanodia}, {Libby-Roberts}, {Mahadevan}, {Ca{\~n}as}, {Gupta}, {Han}, {Kobulnicky}, {Larsen}, {Robertson}, {Rodruck}, {Stefansson}, {Cochran}, {Delamer}, {Diddams}, {Fernandes}, {Halverson}, {Hebb}, {Lin}, {Monson}, {Ninan}, {Roy}, \& {Schwab}}]{Hotnisky2025}
{Hotnisky}, A., {Kanodia}, S., {Libby-Roberts}, J., {et~al.} 2025, \aj, 170, 1, \dodoi{10.3847/1538-3881/add2ef}

\bibitem[{Hotnisky {et~al.}(2025)Hotnisky, Kanodia, Libby-Roberts, Mahadevan, Cañas, Gupta, Han, Kobulnicky, Larsen, Robertson, Rodruck, Stefansson, Cochran, Delamer, Diddams, Fernandes, Halverson, Hebb, Lin, Monson, Ninan, Roy, \& Schwab}]{Hotnisky_2025}
Hotnisky, A., Kanodia, S., Libby-Roberts, J., {et~al.} 2025, The Astronomical Journal, 170, 1, \dodoi{10.3847/1538-3881/add2ef}

\bibitem[{{Howell} {et~al.}(2011){Howell}, {Everett}, {Sherry}, {Horch}, \& {Ciardi}}]{Howell2011}
{Howell}, S.~B., {Everett}, M.~E., {Sherry}, W., {Horch}, E., \& {Ciardi}, D.~R. 2011, \aj, 142, 19, \dodoi{10.1088/0004-6256/142/1/19}

\bibitem[{{Huang} {et~al.}(2020){Huang}, {Vanderburg}, {P{\'a}l}, {Sha}, {Yu}, {Fong}, {Fausnaugh}, {Shporer}, {Guerrero}, {Vanderspek}, \& {Ricker}}]{QLP}
{Huang}, C.~X., {Vanderburg}, A., {P{\'a}l}, A., {et~al.} 2020, Research Notes of the American Astronomical Society, 4, 204, \dodoi{10.3847/2515-5172/abca2e}

\bibitem[{{Ida} \& {Lin}(2005)}]{Ida05}
{Ida}, S., \& {Lin}, D.~N.~C. 2005, \apj, 626, 1045, \dodoi{10.1086/429953}

\bibitem[{{Kagetani} {et~al.}(2023){Kagetani}, {Narita}, {Kimura}, {Hirano}, {Ikoma}, {Ishikawa}, {Giacalone}, {Fukui}, {Kodama}, {Gore}, {Schroeder}, {Hori}, {Kawauchi}, {Watanabe}, {Mori}, {Zou}, {Ikuta}, {Krishnamurthy}, {Zink}, {Hardegree-Ullman}, {Harakawa}, {Kudo}, {Kotani}, {Kurokawa}, {Kusakabe}, {Kuzuhara}, {de Leon}, {Livingston}, {Nishikawa}, {Omiya}, {Palle}, {Parviainen}, {Serizawa}, {Teng}, {Ueda}, \& {Tamura}}]{Kagetani2023}
{Kagetani}, T., {Narita}, N., {Kimura}, T., {et~al.} 2023, \pasj, 75, 713, \dodoi{10.1093/pasj/psad031}

\bibitem[{{Kanodia}(2025)}]{Kanodia2025}
{Kanodia}, S. 2025, \apj, 978, 97, \dodoi{10.3847/1538-4357/ad9823}

\bibitem[{{Kanodia} \& {Wright}(2018)}]{Kanodia18b}
{Kanodia}, S., \& {Wright}, J. 2018, Research Notes of the American Astronomical Society, 2, 4, \dodoi{10.3847/2515-5172/aaa4b7}

\bibitem[{Kanodia {et~al.}(2018)Kanodia, Mahadevan, Ramsey, Stefansson, Monson, Hearty, Blakeslee, Lubar, Bender, Ninan, Sterner, Roy, Halverson, \& Robertson}]{Kanodia18}
Kanodia, S., Mahadevan, S., Ramsey, L.~W., {et~al.} 2018, in Ground-based and Airborne Instrumentation for Astronomy VII, ed. C.~J. Evans, L.~Simard, \& H.~Takami, Vol. 10702, International Society for Optics and Photonics (SPIE), 107026Q, \dodoi{10.1117/12.2313491}

\bibitem[{{Kanodia} {et~al.}(2024{\natexlab{a}}){Kanodia}, {Ca{\~n}as}, {Mahadevan}, {Ford}, {Helled}, {Anderson}, {Boss}, {Cochran}, {Delamer}, {Han}, {Libby-Roberts}, {Lin}, {M{\"u}ller}, {Robertson}, {Stef{\'a}nsson}, \& {Teske}}]{motive}
{Kanodia}, S., {Ca{\~n}as}, C.~I., {Mahadevan}, S., {et~al.} 2024{\natexlab{a}}, \aj, 167, 161, \dodoi{10.3847/1538-3881/ad27cb}

\bibitem[{{Kanodia} {et~al.}(2024{\natexlab{b}}){Kanodia}, {Gupta}, {Ca{\~n}as}, {Bernab{\`o}}, {Reji}, {Han}, {Brady}, {Seifahrt}, {Cochran}, {Morrell}, {Basant}, {Bean}, {Bender}, {de Beurs}, {Bieryla}, {Birkholz}, {Brown}, {Chapman}, {Ciardi}, {Clark}, {Cotter}, {Diddams}, {Halverson}, {Hawley}, {Hebb}, {Holcomb}, {Howell}, {Kobulnicky}, {Kowalski}, {Larsen}, {Libby-Roberts}, {Lin}, {Lund}, {Luque}, {Monson}, {Ninan}, {Parker}, {Patel}, {Rodruck}, {Ross}, {Roy}, {Schwab}, {Stef{\'a}nsson}, {Thoms}, \& {Vanderburg}}]{KanodiaSixGEMS}
{Kanodia}, S., {Gupta}, A.~F., {Ca{\~n}as}, C.~I., {et~al.} 2024{\natexlab{b}}, \aj, 168, 235, \dodoi{10.3847/1538-3881/ad7796}

\bibitem[{{Kanodia} {et~al.}(2025){Kanodia}, {Ca{\~n}as}, {Mahadevan}, {Lin}, {Kobulnicky}, {Karfs}, {Birkholz}, {Gupta}, {Everett}, {Rodruck}, {Glusman}, {Han}, {Cochran}, {Bender}, {Diddams}, {Krolikowski}, {Halverson}, {Libby-Roberts}, {Ninan}, {Robertson}, {Roy}, {Schwab}, \& {Stef{\'a}nsson}}]{Kanodia2025b}
{Kanodia}, S., {Ca{\~n}as}, C.~I., {Mahadevan}, S., {et~al.} 2025, arXiv e-prints, arXiv:2506.17861, \dodoi{10.48550/arXiv.2506.17861}

\bibitem[{Kasper {et~al.}(2016)Kasper, Ellis, Yeigh, Kobulnicky, Jang-Condell, Kelley, Bucher, \& Weger}]{Kasper16}
Kasper, D.~H., Ellis, T.~G., Yeigh, R.~R., {et~al.} 2016, Publications of the Astronomical Society of the Pacific, 128, 105005, \dodoi{10.1088/1538-3873/128/968/105005}

\bibitem[{{Kipping}(2013)}]{Kipping2013}
{Kipping}, D.~M. 2013, \mnras, 435, 2152, \dodoi{10.1093/mnras/stt1435}

\bibitem[{{Kunimoto} {et~al.}(2022{\natexlab{a}}){Kunimoto}, {Tey}, {Fong}, {Hesse}, {Shporer}, {Fausnaugh}, {Vanderspek}, \& {Ricker}}]{Kunimoto22}
{Kunimoto}, M., {Tey}, E., {Fong}, W., {et~al.} 2022{\natexlab{a}}, Research Notes of the American Astronomical Society, 6, 236, \dodoi{10.3847/2515-5172/aca158}

\bibitem[{{Kunimoto} {et~al.}(2022{\natexlab{b}}){Kunimoto}, {Daylan}, {Guerrero}, {Fong}, {Bryson}, {Ricker}, {Fausnaugh}, {Huang}, {Sha}, {Shporer}, {Vanderburg}, {Vanderspek}, \& {Yu}}]{faint-star}
{Kunimoto}, M., {Daylan}, T., {Guerrero}, N., {et~al.} 2022{\natexlab{b}}, \apjs, 259, 33, \dodoi{10.3847/1538-4365/ac5688}

\bibitem[{Laughlin {et~al.}(2004)Laughlin, Bodenheimer, \& Adams}]{Laughlin}
Laughlin, G., Bodenheimer, P., \& Adams, F.~C. 2004, 612, L73, \dodoi{10.1086/424384}

\bibitem[{Mahadevan {et~al.}(2012)Mahadevan, Ramsey, Bender, Terrien, Wright, Halverson, Hearty, Nelson, Burton, Redman, Osterman, Diddams, Kasting, Endl, \& Deshpande}]{Mahadevan12}
Mahadevan, S., Ramsey, L., Bender, C., {et~al.} 2012, in Ground-based and Airborne Instrumentation for Astronomy IV, ed. I.~S. McLean, S.~K. Ramsay, \& H.~Takami, Vol. 8446, International Society for Optics and Photonics (SPIE), 84461S, \dodoi{10.1117/12.926102}

\bibitem[{Mahadevan {et~al.}(2014)Mahadevan, Ramsey, Terrien, Halverson, Roy, Hearty, Levi, Stefansson, Robertson, Bender, Schwab, \& Nelson}]{Mahadevan14}
Mahadevan, S., Ramsey, L.~W., Terrien, R., {et~al.} 2014, in Ground-based and Airborne Instrumentation for Astronomy V, ed. S.~K. Ramsay, I.~S. McLean, \& H.~Takami, Vol. 9147, International Society for Optics and Photonics (SPIE), 91471G, \dodoi{10.1117/12.2056417}

\bibitem[{{NASA Exoplanet Archive}(2025)}]{NASA_EXOARCH}
{NASA Exoplanet Archive}. 2025, Planetary Systems Composite Parameters, Version: 2025-04-15 00:00,  NExScI-Caltech/IPAC, \dodoi{10.26133/NEA13}

\bibitem[{{Ninan} {et~al.}(2018){Ninan}, {Bender}, {Mahadevan}, {Ford}, {Monson}, {Kaplan}, {Terrien}, {Roy}, {Robertson}, {Kanodia}, \& {Stefansson}}]{HxRGproc}
{Ninan}, J.~P., {Bender}, C.~F., {Mahadevan}, S., {et~al.} 2018, in Society of Photo-Optical Instrumentation Engineers (SPIE) Conference Series, Vol. 10709, High Energy, Optical, and Infrared Detectors for Astronomy VIII, ed. A.~D. {Holland} \& J.~{Beletic}, 107092U, \dodoi{10.1117/12.2312787}

\bibitem[{{Osborn} \& {Bayliss}(2020)}]{Osborn2020}
{Osborn}, A., \& {Bayliss}, D. 2020, \mnras, 491, 4481, \dodoi{10.1093/mnras/stz3207}

\bibitem[{{Pfalzner} \& {Dincer}(2024)}]{Pfalzner2024}
{Pfalzner}, S., \& {Dincer}, F. 2024, \apj, 963, 122, \dodoi{10.3847/1538-4357/ad1bef}

\bibitem[{Ramsey {et~al.}(1998)Ramsey, Adams, III, Booth, Cornell, Fowler, Gaffney, Glaspey, Good, Hill, Kelton, Krabbendam, Long, MacQueen, Ray, Ricklefs, Sage, Sebring, Spiesman, \& Steiner}]{Ramsey98}
Ramsey, L.~W., Adams, M.~T., III, T. G.~B., {et~al.} 1998, in Advanced Technology Optical/IR Telescopes VI, ed. L.~M. Stepp, Vol. 3352, International Society for Optics and Photonics (SPIE), 34 -- 42, \dodoi{10.1117/12.319287}

\bibitem[{{Rasio} \& {Ford}(1996)}]{Rasio1996}
{Rasio}, F.~A., \& {Ford}, E.~B. 1996, Science, 274, 954, \dodoi{10.1126/science.274.5289.954}

\bibitem[{Ricker {et~al.}(2014)Ricker, Winn, Vanderspek, Latham, Bakos, Bean, Berta-Thompson, Brown, Buchhave, Butler, Butler, Chaplin, Charbonneau, Christensen-Dalsgaard, Clampin, Deming, Doty, Lee, Dressing, Dunham, Endl, Fressin, Ge, Henning, Holman, Howard, Ida, Jenkins, Jernigan, Johnson, Kaltenegger, Kawai, Kjeldsen, Laughlin, Levine, Lin, Lissauer, MacQueen, Marcy, McCullough, Morton, Narita, Paegert, Palle, Pepe, Pepper, Quirrenbach, Rinehart, Sasselov, Sato, Seager, Sozzetti, Stassun, Sullivan, Szentgyorgyi, Torres, Udry, \& Villasenor}]{Ricker15}
Ricker, G.~R., Winn, J.~N., Vanderspek, R., {et~al.} 2014, Journal of Astronomical Telescopes, Instruments, and Systems, 1, 014003, \dodoi{10.1117/1.JATIS.1.1.014003}

\bibitem[{{Salvatier} {et~al.}(2015){Salvatier}, {Wiecki}, \& {Fonnesbeck}}]{pymc3_Salvatier}
{Salvatier}, J., {Wiecki}, T., \& {Fonnesbeck}, C. 2015, arXiv e-prints, arXiv:1507.08050, \dodoi{10.48550/arXiv.1507.08050}

\bibitem[{{Scott} {et~al.}(2018){Scott}, {Howell}, {Horch}, \& {Everett}}]{Scott2018}
{Scott}, N.~J., {Howell}, S.~B., {Horch}, E.~P., \& {Everett}, M.~E. 2018, \pasp, 130, 054502, \dodoi{10.1088/1538-3873/aab484}

\bibitem[{{Srinath} {et~al.}(2014){Srinath}, {McGurk}, {Rockosi}, {Kupke}, {Gavel}, {Cabak}, {Cowley}, {Peck}, {Ratliff}, {Gates}, {Dillon}, {Norton}, \& {Reining}}]{srinath_swimming_2014}
{Srinath}, S., {McGurk}, R., {Rockosi}, C., {et~al.} 2014, in Society of Photo-Optical Instrumentation Engineers (SPIE) Conference Series, Vol. 9148, Adaptive Optics Systems IV, ed. E.~{Marchetti}, L.~M. {Close}, \& J.-P. {Vran}, 91482Z, \dodoi{10.1117/12.2055672}

\bibitem[{{Stassun} {et~al.}(2019){Stassun}, {Oelkers}, {Paegert}, {Torres}, {Pepper}, {De Lee}, {Collins}, {Latham}, {Muirhead}, {Chittidi}, {Rojas-Ayala}, {Fleming}, {Rose}, {Tenenbaum}, {Ting}, {Kane}, {Barclay}, {Bean}, {Brassuer}, {Charbonneau}, {Ge}, {Lissauer}, {Mann}, {McLean}, {Mullally}, {Narita}, {Plavchan}, {Ricker}, {Sasselov}, {Seager}, {Sharma}, {Shiao}, {Sozzetti}, {Stello}, {Vanderspek}, {Wallace}, \& {Winn}}]{TESS_Catalog}
{Stassun}, K.~G., {Oelkers}, R.~J., {Paegert}, M., {et~al.} 2019, \aj, 158, 138, \dodoi{10.3847/1538-3881/ab3467}

\bibitem[{{Stefansson} {et~al.}(2016){Stefansson}, {Hearty}, {Robertson}, {Mahadevan}, {Anderson}, {Levi}, {Bender}, {Nelson}, {Monson}, {Blank}, {Halverson}, {Henderson}, {Ramsey}, {Roy}, {Schwab}, \& {Terrien}}]{Stefansson2016}
{Stefansson}, G., {Hearty}, F., {Robertson}, P., {et~al.} 2016, \apj, 833, 175, \dodoi{10.3847/1538-4357/833/2/175}

\bibitem[{{Stef\'ansson} {et~al.}(2020){Stef\'ansson}, {Ca{\~n}as}, {Wisniewski}, {Robertson}, {Mahadevan}, {Maney}, {Kanodia}, {Beard}, {Bender}, {Brunt}, {Clemens}, {Cochran}, {Diddams}, {Endl}, {Ford}, {Fredrick}, {Halverson}, {Hearty}, {Hebb}, {Huehnerhoff}, {Jennings}, {Kaplan}, {Levi}, {Lubar}, {Metcalf}, {Monson}, {Morris}, {Ninan}, {Nitroy}, {Ramsey}, {Roy}, {Schwab}, {Sigurdsson}, {Terrien}, \& {Wright}}]{Stefansson2020}
{Stef\'ansson}, G., {Ca{\~n}as}, C., {Wisniewski}, J., {et~al.} 2020, \aj, 159, 100, \dodoi{10.3847/1538-3881/ab5f15}

\bibitem[{{Stef{\'a}nsson} {et~al.}(2023){Stef{\'a}nsson}, {Mahadevan}, {Miguel}, {Robertson}, {Delamer}, {Kanodia}, {Ca{\~n}as}, {Winn}, {Ninan}, {Terrien}, {Holcomb}, {Ford}, {Zawadzki}, {Bowler}, {Bender}, {Cochran}, {Diddams}, {Endl}, {Fredrick}, {Halverson}, {Hearty}, {Hill}, {Lin}, {Metcalf}, {Monson}, {Ramsey}, {Roy}, {Schwab}, {Wright}, \& {Zeimann}}]{Stefansson2023}
{Stef{\'a}nsson}, G., {Mahadevan}, S., {Miguel}, Y., {et~al.} 2023, Science, 382, 1031, \dodoi{10.1126/science.abo0233}

\bibitem[{{Wright} \& {Eastman}(2014)}]{Wright14}
{Wright}, J.~T., \& {Eastman}, J.~D. 2014, \pasp, 126, 838, \dodoi{10.1086/678541}

\bibitem[{{Yee} {et~al.}(2017){Yee}, {Petigura}, \& {von Braun}}]{Yee2017}
{Yee}, S.~W., {Petigura}, E.~A., \& {von Braun}, K. 2017, \apj, 836, 77, \dodoi{10.3847/1538-4357/836/1/77}

\bibitem[{{Zechmeister} {et~al.}(2018){Zechmeister}, {Reiners}, {Amado}, {Azzaro}, {Bauer}, {B{\'e}jar}, {Caballero}, {Guenther}, {Hagen}, {Jeffers}, {Kaminski}, {K{\"u}rster}, {Launhardt}, {Montes}, {Morales}, {Quirrenbach}, {Reffert}, {Ribas}, {Seifert}, {Tal-Or}, \& {Wolthoff}}]{Zechmeister18}
{Zechmeister}, M., {Reiners}, A., {Amado}, P.~J., {et~al.} 2018, \aap, 609, A12, \dodoi{10.1051/0004-6361/201731483}

\end{thebibliography}
\bibliographystyle{aasjournal}



\end{document}